\documentclass[12pt,a4paper]{article}
\usepackage{a4wide}
\usepackage{epsfig}
\usepackage{cite}
\usepackage{amsmath}
\usepackage{graphicx}
\usepackage{latexsym}
\usepackage{array}
\usepackage{multirow}

\newlength{\absize}
\newcommand{\dd}{\mbox{{\rm d}}}

\def\lsim{\mathrel{\rlap{\raise 2.5pt \hbox{$<$}}\lower 2.5pt
\hbox{$\sim$}}}
\def\gsim{\mathrel{\rlap{\raise 2.5pt \hbox{$>$}}\lower 2.5pt
\hbox{$\sim$}}}

\newcommand{\Lumint}{{\cal L}_{\rm int}}

\begin{document}

\thispagestyle{empty}
\renewcommand{\thefootnote}{\fnsymbol{footnote}}
\newpage\normalsize
\pagestyle{plain}
\setlength{\baselineskip}{4ex}\par
\setcounter{footnote}{0}
\renewcommand{\thefootnote}{\arabic{footnote}}
\newcommand{\preprint}[1]{%
\begin{flushright}
\setlength{\baselineskip}{3ex} #1
\end{flushright}}
\renewcommand{\title}[1]{%
\begin{center}
\LARGE #1
\end{center}\par}
\renewcommand{\author}[1]{%
\vspace{2ex}
{\Large
\begin{center}
  \setlength{\baselineskip}{3ex} #1 \par
\end{center}}}
\renewcommand{\thanks}[1]{\footnote{#1}}
\renewcommand{\abstract}[1]{%
\vspace{2ex}
\normalsize
\begin{center}
 \centerline{\bf Abstract}\par
 \vspace{2ex}
 \parbox{\absize}{#1\setlength{\baselineskip}{2.5ex}\par}
\end{center}}

\vspace*{4mm} 

\title{Identification of extra neutral gauge bosons at the International
Linear Collider} \vfill

\author{P. Osland,$^{a,}$\footnote{E-mail: per.osland@ift.uib.no}
A. A. Pankov,$^{b,}$\footnote{E-mail: pankov@ictp.it}
and A. V. Tsytrinov$^{b,}$\footnote{E-mail: tsytrin@rambler.ru}}

\begin{center}
$^{a}$Department of Physics and Technology, University of Bergen,
Postboks 7803, N-5020  Bergen, Norway\\
$^{b}$The Abdus Salam ICTP Affiliated Centre, Technical
University of Gomel, 246746 Gomel, Belarus\
\end{center}
%
%
%
\vfill

\abstract{Heavy neutral gauge bosons, $Z^\prime$s, are predicted by
many theoretical schemes of physics beyond the Standard Model, and intensive
searches for their signatures will be performed at present and future
high energy colliders.
It is quite possible that $Z^\prime$s are heavy enough to lie beyond
the discovery reach expected at the CERN Large Hadron Collider LHC, in
which case only indirect signatures of $Z^\prime$ exchanges
may occur at future colliders, through deviations of the measured cross
sections from the Standard Model predictions. We here discuss in
this context the foreseeable sensitivity to $Z^\prime$s of
fermion-pair production cross sections at an
$e^+e^-$ linear collider, especially as regards the potential
of distinguishing different $Z^\prime$ models once such deviations
are observed. Specifically, we assess the discovery and
identification reaches on $Z^\prime$ gauge bosons pertinent
to the $E_6$, LR, ALR and SSM classes of models, that should be attained
at the planned International Linear Collider (ILC).
With the high experimental accuracies expected at the ILC, the
discovery and the identification reaches
on the $Z^\prime$ models under consideration could be increased
substantially. In particular, the identification among
the different models could be
achieved for values of $Z^\prime$ masses in the discovery (but beyond
the identification) reach of the LHC. An important role in enhancing
such reaches is played by the  electron (and possibly the positron)
longitudinally polarized beams. Also, although the purely leptonic
processes are experimentally cleaner, the measurements of $c$- and
$b$-quark pair production cross sections are found to carry
important, and complementary, information on these searches.}

\vspace*{20mm} \setcounter{footnote}{0} \vfill

\newpage
\setcounter{footnote}{0}
\renewcommand{\thefootnote}{\arabic{footnote}}

\section{Introduction} \label{sect:introduction}

Electroweak theories beyond the Standard Model (SM) based on
spontaneously broken extended gauge symmetries naturally
envisage the existence of heavy, neutral, vector bosons $Z^\prime$.
The variety of the proposed $Z^\prime$ models is somewhat broad,
and for definiteness in the sequel we shall focus on the so-called
$Z^\prime_{\rm SSM}$, $Z^\prime_{E_6}$, $Z^\prime_{\rm LR}$ and
$Z^\prime_{\rm ALR}$ models. Particular attention has recently been
devoted to the phenomenological properties and the search reaches
on such scenarios, and in some sense we may consider these $Z^\prime$
models as representative of this New Physics (NP)
sector \cite{resonances,Erler:2009jh}.

\par
A typical manifestation of the production of such states
is represented by (narrow) peaks observed in the cross sections
for processes among SM particles at high energy accelerators, for
example in the invariant mass distributions for Drell-Yan dilepton
pair production at the Fermilab Tevatron or at the CERN LHC
hadronic colliders. Current experimental search limits on
$M_{Z^\prime}$ at 95\% C.L., from Drell-Yan cross sections at
the Tevatron, generally range in the interval 0.8--1 TeV,
depending on the particular $Z^\prime$ model being
tested \cite{Tev:2007sb}. Even higher 95\% C.L. limits, of
the order of 1.14--1.4 TeV are obtained for the $Z^\prime_\chi$,
$Z^\prime_{\rm LR}$, and $Z^\prime_{\rm SSM}$ models, from
electroweak high precision data \cite{Erler:2009jh}.

\par
Clearly, the eventual discovery of a peak should be supplemented by
the verification of the spin-1 of the assumed underlying $Z^\prime$,
vs. the alternative spin-2 and spin-0 hypotheses corresponding, e.g.,
to exchanges of a Randall-Sundrum graviton
resonance~\cite{Randall:1999ee} or a
sneutrino~\cite{Kalinowski:1997bc}. This kind of analysis relies on
appropriate angular differential distributions and/or angular
asymmetries. Finally, once the spin-1 has been established, the
particular $Z^\prime$ scenario pertinent to the observed signal should
be identified, see, e.g.,
Refs.~\cite{Allanach:2000nr,Dittmar:2003ir,Cousins:2005pq,Godfrey:2005pm,
  Feldman:2006wb,Petriello:2008zr,Osland:2008sy,Murayama:2009jz,
  Osland:2009tn,Salvioni:2009mt}. From studies of Drell-Yan processes
at the LHC with a time-integrated luminosity of 100~fb$^{-1}$, it
turns out that one can expect, at the 5-$\sigma$ level, discovery
limits on $M_{Z^\prime}$ of the order of 4--4.5 TeV, spin-1
identification up to $M_{Z^\prime}\simeq 2.5$--3~TeV and potential of
distinction among the individual $Z^\prime$ models up to
$M_{Z^\prime}\simeq 2.1$~TeV (95\% C.L.).

\par
For masses above the direct search limits mentioned above, and LHC
luminosity at the design value, access to $Z^\prime$
manifestations may be provided by indirect, virtual exchange
effects causing deviations of cross sections from the SM predictions,
if $M_{Z^\prime}$ is not excessively heavy. However, at the LHC,
model identification from Drell-Yan dilepton mass distributions and
forward-backward asymmetries may be problematic due to limited statistics
\cite{Rizzo:2009pu}.

\par
An alternative resource for the observation of virtual heavy
gauge boson exchanges should be represented by the next
generation $e^+e^-$ International Linear Collider (ILC), with
center of mass energy $\sqrt s=0.5$--1~TeV and typical
time-integrated luminosities ${\cal L}_{\rm int}\sim
0.5$--1~ab$^{-1}$~\cite{:2007sg,Djouadi:2007ik}, and the really
high precision measurements that will be possible there. Indeed,
the baseline configuration envisages a very high electron beam
polarization (larger than 80\%). Also positron beam
polarization, around 30\%, might be initially obtainable and
perhaps already available for
physics. This polarization could be raised to about 60\% or higher
in the ultimate upgrade of the machine. The polarization option
might represent an asset in order to enhance the discovery reaches
and identification sensitivities on NP models of any
kind \cite{MoortgatPick:2005cw},
therefore also on $Z^\prime$ exchanges in interactions of SM particles.
Previous analyses, based on
various final state channels and possible experimental
observables, show that sensitivities to quite high
$Z^\prime$ masses could in principle be attained at the ILC (qualitatively,
of the order of $M_{Z'}\sim (10-20)\cdot \sqrt s$ for the highest
planned luminosity, see, e.g.,
\cite{resonances,Cvetic:1995zs,Riemann:2001bb,
Djouadi:1991sx,Gulov:2004sg,Pankov:2005kd}, and references therein). 
The ILC parameters have recently been fixed in the Reference Design
Report \cite{:2007sg}, so that it should be interesting to
reconsider the identification of $Z^\prime$ models in the light of
the numbers reported there.

\par
We will here focus on the fermion-antifermion production reactions
at the polarized ILC:
\begin{equation}
e^++e^-\to f + {\bar f},\qquad\quad f=e,\mu,\tau,c,b.
\label{proc_ff}
\end{equation}
As basic experimental observables for the $Z^\prime$ analysis, as
an alternative to integrated observables like the total cross
sections and/or angular-integrated asymmetries, we here choose
the differential angular distributions for the above
processes, that allow to exploit the information contained in the
different portions of the final state phase space by a binned
analysis. 
Particular emphasis will be given to the comparison between the
cases of unpolarized and polarized initial beams, as regards the
expected potential of ILC in identifying the $Z^\prime$ models of
interest here, for $M_{Z^\prime}$ values of the order of and
beyond the limits accessible at the LHC.

\par
Indeed, concerning the $Z^\prime$ mass, there are two scenarios. The first
one is represented by the interval in $M_{Z^\prime}$ between the expected
identification and discovery limits at the LHC: here, we can assume the
$Z^\prime$ to have already been discovered at some $M_{Z^\prime}$ (but
the model not identified), so that the model identification (or
equivalently the determination of the coupling constants) could be
performed at the ILC, based on the deviations of cross sections
from the SM predictions for the determined $Z^\prime$ mass. For
earlier attempts along this line see, e.g., Ref.~\cite{Weiglein:2004hn}.
The second mass range is above the LHC discovery limit and, here, with
$M_{Z^\prime}$ unknown, both discovery and identification reaches should
be assessed for the ILC.

\par
In the following, in Sec.~\ref{sect:observ} we give a brief
introduction to the different $Z^\prime$ models considered in the
analysis, and give the corresponding leading order expressions of the
polarized differential cross sections for processes (\ref{proc_ff}),
mostly in order to establish the notations. In
Secs.~\ref{sect:discovery} and \ref{sect:distinction} we present the
results of our analysis for the discovery and identification reaches
on the individual $Z^\prime$ models at the ILC; and finally,
Sec.~\ref{sect:concl} contains some concluding remarks.

\section{Polarized observables and $Z'$ models}
\label{sect:observ}
The analysis at the ILC is somewhat different from the corresponding
studies of Drell-Yan processes at the LHC. Deviations of the
various observables from SM predictions, such as cross sections
and asymmetries, due to the interference of the SM amplitude with
$s$-channel exchanges of the $Z^\prime$, graviton resonance $G$ or
sneutrino $\tilde\nu$, might be observed at the ILC. However, in the
latter case, there is no interference of $\tilde\nu$ with the SM
exchanges \cite{Kalinowski:1997bc}. Conversely, in the case of the
spin-2 KK graviton exchange, the interference with the SM
exchanges vanishes when one integrates over the full angular range
\cite{Hewett:1998sn}, whereas for differential observables such
interference survives. Nevertheless, it turns out
\cite{Davoudiasl:2000jd} that the sensitivity at the ILC with
$\sqrt{s}=0.5$ TeV and $\Lumint=500$ fb$^{-1}$ to a KK graviton
resonance in processes (\ref{proc_ff}) is of the order 0.8~TeV
(1.9~TeV) for the graviton coupling constant $c=0.01$ ($c=0.1$),
which is well within the expectations for discovery {\it and}
identification at the LHC. Accordingly, the KK excitation would
have been either discovered or excluded by the time the ILC will
be operating. Therefore, from the considerations above, one can
conclude that for $\sqrt{s}<M_{Z'}$, as will be the case for the ILC,
only the interference of the $Z^\prime$ amplitude with the SM one
could be visible at the ILC in processes (\ref{proc_ff}).
Accordingly, it might not be so indispensible to
perform angular analyses such as those foreseen for Drell-Yan
processes at the LHC, in order to differentiate the spin of the
exchanged intermediate heavy quantum states, because only $Z^\prime$
should be able to lead to appreciable interference
effects.

\par
The polarized differential cross section for the Bhabha process
$e^++e^-\to e^++e^-$, where $\gamma$ and $Z$ can be exchanged also
in the $t$-channel, can be written at leading order as (see, e.g.,
Refs.~\cite{Schrempp:1987zy,Pankov:2002qk}):
\begin{eqnarray}
\frac{\dd\sigma(P^-,P^+)}{\dd z}
&=&\frac{(1+P^-)\,(1-P^+)}4\,\frac{\dd\sigma_{\rm R}}{\dd z}+
\frac{(1-P^-)\,(1+P^+)}4\,\frac{\dd\sigma_{\rm L}}{\dd z}
\nonumber \\
&+&\frac{(1+P^-)\,(1+P^+)}4\,\frac{\dd\sigma_{{\rm RL},t}}{\dd z}+
\frac{(1-P^-)\,(1-P^+)}4 \,\frac{\dd\sigma_{{\rm LR},t}}{\dd z},
\label{cross}
\end{eqnarray}
with the decomposition
\begin{equation}
\frac{\dd\sigma_{\rm L}}{\dd z}= \frac{\dd\sigma_{{\rm LL}}}{\dd z}
+ \frac{\dd\sigma_{{\rm LR},s}}{\dd z}, \qquad
\frac{\dd\sigma_{\rm R}}{\dd z}= \frac{\dd\sigma_{{\rm RR}}}{\dd z}
+ \frac{\dd\sigma_{{\rm RL},s}}{\dd z}. \label{sigP}
\end{equation}
In Eqs.~(\ref{cross}) and (\ref{sigP}), the subscripts $t$ and $s$
denote helicity cross sections with SM $\gamma$ and $Z$ exchanges
in the corresponding channels, $z=\cos\theta$ and the subscripts
$\rm L$, $\rm R$ denote the respective helicities, $P^-$ and $P^+$
denote the degrees of longitudinal polarization of the $e^-$ and
$e^+$ beams, respectively.\footnote{In the recent review
\cite{MoortgatPick:2005cw}, the opposite sign convention for positron
polarization was adopted.} In terms of helicity amplitudes:
\begin{alignat}{2}
\frac{\dd\sigma_{{\rm LL}}}{\dd z} &= \frac{2\pi\alpha_{\rm
e.m.}^2}{s}\,\big\vert G^{ee}_{{\rm LL},s} + G^{ee}_{{\rm LL},t}
\big\vert^2, &\quad\ \ \frac{\dd\sigma_{{\rm RR}}}{\dd z} &=
\frac{2\pi\alpha_{\rm e.m.}^2}{s}\, \big\vert G^{ee}_{{\rm
RR},s}+G^{ee}_{{\rm RR},t}\big\vert^2 ,
\nonumber \\
\frac{\dd\sigma_{{\rm LR},t}}{\dd z} &=\frac{\dd\sigma_{{\rm
RL},t}}{\dd z}= \frac{2\pi\alpha_{\rm e.m.}^2}{s}\, \big\vert
G^{ee}_{{\rm LR},t}\big\vert^2, &\quad \frac{\dd\sigma_{{\rm
LR},s}}{\dd z} &= \frac{\dd\sigma_{{\rm RL},s}}{\dd z}=
\frac{2\pi\alpha_{\rm e.m.}^2}{s}\,\big\vert G^{ee}_{{\rm
LR},s}\big\vert^2. \label{helsig}
\end{alignat}

\par
According to the previous considerations the amplitudes
$G^{ee}_{\alpha\beta,i}$, with $\alpha,\beta={\rm L,R}$ and
$i=s,t$, are given by the sum of the SM $\gamma, Z$ exchanges plus
deviations representing the effect induced by a $Z^\prime$ boson:
\begin{alignat}{2}
G^{ee}_{{\rm LL},s} &={u}\,\left(\frac{1}{s}+\frac{(g^e_{\rm L})^2}
{s-M^2_Z}+ \frac{(g^e_{\rm L}{}^\prime)^2}{s-M^2_{Z'}} \right), &\quad
G^{ee}_{{\rm LL},t}&={u}\,\left(\frac{1}{t}+\frac{(g^e_{\rm L})^2}
{t-M^2_Z} +\frac{(g^e_{\rm L}{}^\prime)^2}{t-M^2_{Z'}} \right),
\nonumber \\
G^{ee}_{{\rm RR},s}&={u}\,\left(\frac{1}{s}+ \frac{(g^e_{\rm R})^2}
{s-M^2_Z} + \frac{(g^e_{\rm R}{}^\prime)^2}{s-M^2_{Z'}}\right), &\quad
G^{ee}_{{\rm RR},t}&={u}\,\left(\frac{1}{t}+\frac{(g^e_{\rm R})^2}
{t-M^2_Z} +\frac{(g^e_{\rm R}{}^\prime)^2}{t-M^2_{Z'}}\right),
\nonumber \\
G^{ee}_{{\rm LR},s}&={t}\,\left(\frac{1}{s}+\frac{g^e_{\rm R}\hskip
2pt g^e_{\rm L}}{s-M^2_Z}+ \frac{g^e_{\rm R}{}^\prime\hskip 2pt
g^e_{\rm L}{}^\prime}{s-M^2_{Z'}}\right), &\qquad
G^{ee}_{{\rm LR},t}&=s\,\left(\frac{1}{t}+\frac{g^e_{\rm R}\hskip 2pt
g^e_{\rm L}}{t-M^2_Z}+\frac{g^e_{\rm R}{}^\prime\hskip 2pt
g^e_{\rm L}{}^\prime}{t-M^2_{Z'}}\right). \label{helamp}
\end{alignat}
Here, $u,t=-s(1\pm z)/2$ (we are neglecting fermion masses),
$g_{\rm L}=-\cot{2\,\theta_W}$ and
$g_{\rm R}=\tan\theta_W$ with $\theta_W$ the electroweak mixing
angle, whereas $g'_{\rm L}$ and $g'_{\rm R}$ are characteristic
of the particular $Z^\prime$ model.
In the annihilation channel, below the $Z^\prime$ mass, the $Z^\prime$
interference with the SM will be destructive in the LL and RR cross
sections, whereas it can be of either sign in the LR and RL cross
sections.

\par
The polarized differential cross section for the leptonic channels
$e^+e^-\to l^+l^-$ with $l=\mu,\tau$ can be obtained directly from
Eq.~(\ref{cross}), basically by dropping the $t$-channel
contributions. The same is true, after some obvious substitutions,
for the annihilations into $c{\bar c}$ and $b{\bar b}$ final
states, in which case also the color ($N_C$) and QCD correction
factors, $C_s\simeq N_C\,[1+\alpha_s/\pi+1.4\,(\alpha_s/\pi)^2]$,
must be taken into account. The $s$-channel helicity amplitudes
for the process (\ref{proc_ff}) with $f\ne e,t$
 can be written as \cite{Schrempp:1987zy}:
\begin{alignat}{2}
G^{ef}_{\alpha\alpha,s}
&={u}\,\left(\frac{Q_e\,Q_f}{s}+\frac{g^e_{\alpha}g^f_{\alpha}}{s-M^2_Z}+
\frac{g'^e_{\alpha}g'^f_{\alpha}}{s-M^2_{Z'}} \right), &\quad
G^{ef}_{\alpha\beta,s}
&={t}\,\left(\frac{Q_e\,Q_f}{s}+\frac{g^e_{\alpha}g^f_{\beta}}{s-M^2_Z}+
\frac{g'^e_{\alpha}g'^f_{\beta}}{s-M^2_{Z'}} \right),
\label{helamp_ff}
\end{alignat}
where in the latter expression $\alpha\ne\beta$.

\par
As anticipated, the $Z^\prime$ models that will be considered
in our analysis are the following:
\begin{itemize}
\item[(i)] The $Z'$ scenarios originating from the
exceptional group $E_{6}$ spontaneous breaking are defined
in terms of a mixing angle $\beta$. The specific values $\beta=0$,
$\beta=\pi/2$ and $\beta=-\arctan{\sqrt{5/3}}$, correspond to
different $E_{6}$ breaking patterns and define the popular
scenarios $Z^\prime_\chi$, $Z^\prime_\psi$ and $Z^\prime_\eta$,
respectively.

\item[(ii)] The left-right models, originating from the
breaking of an $SO(10)$ grand-unification symmetry, and where
the corresponding $Z^\prime_{\rm LR}$ couples to a combination
of right-handed and $B-L$ neutral currents ($B$ and $L$ denote
baryon and lepton currents), specified by a real parameter
$\alpha_{\rm LR}$ bounded by
$\sqrt{2/3} \lsim \alpha_{\rm LR}\lsim\sqrt{2}$.
The particular value $\alpha_{\rm LR}=\sqrt 2$ corresponds to
a pure L-R symmetric model (LRS).

\item[(iii)]
The $Z^\prime_{\rm ALR}$ predicted by the `alternative' left-right
scenario.

\item[(iv)]
The so-called sequential $Z^\prime_\text{SSM}$, where the
couplings to fermions are the same as those of the SM $Z$.
\end{itemize}
Detailed descriptions of these models, as well as the specific
references, can be found, e.~g., in Ref.~\cite{resonances}. All
numerical values of the $Z^\prime$ couplings needed in
Eq.~(\ref{helamp}) are collected, for example, in
Table~1 of Ref.~\cite{Osland:2009tn}.
\section{Discovery of $Z'$}
\label{sect:discovery}

In the absence of available data, the assessment of
the expected `discovery reaches' on the various $Z^\prime$s
needs the definition of a `distance' between the NP model
predictions and those of the SM for the basic observables
that will be measured.
The former predictions parametrically depend on the $Z^\prime$ mass
and its corresponding coupling constants, while the latter
ones are calculated using the parameters known from the SM fits.
Such a comparison can be performed by a standard $\chi^2$-like
procedure. As anticipated in Sec.~1, we divide the full angular
range into bins and identify the basic observables with the polarized
differential angular distributions for processes (\ref{proc_ff}),
${\cal O}={\rm d}\sigma(P^-,P^+)/{\rm d} z$, in each bin.
Correspondingly, the relevant $\chi^2$ can symbolically be
defined as:
\begin{equation}
\chi^2({\cal O})=\sum_{f}\sum_{\{P^-,\ P^+\}} \sum_{\rm
bins}\frac{\left[{\cal O}({\rm SM}+Z^\prime) - {\cal O}({\rm
SM})\right]_{\rm bin}^2 }{\left(\delta{\cal O}_{\rm
bin}\right)^2}. \label{chisquare}
\end{equation}

\begin{figure}[htbp] %
\vspace*{0.0cm} \centerline{\hspace*{-1.8cm}
\includegraphics[width=16cm,angle=0]{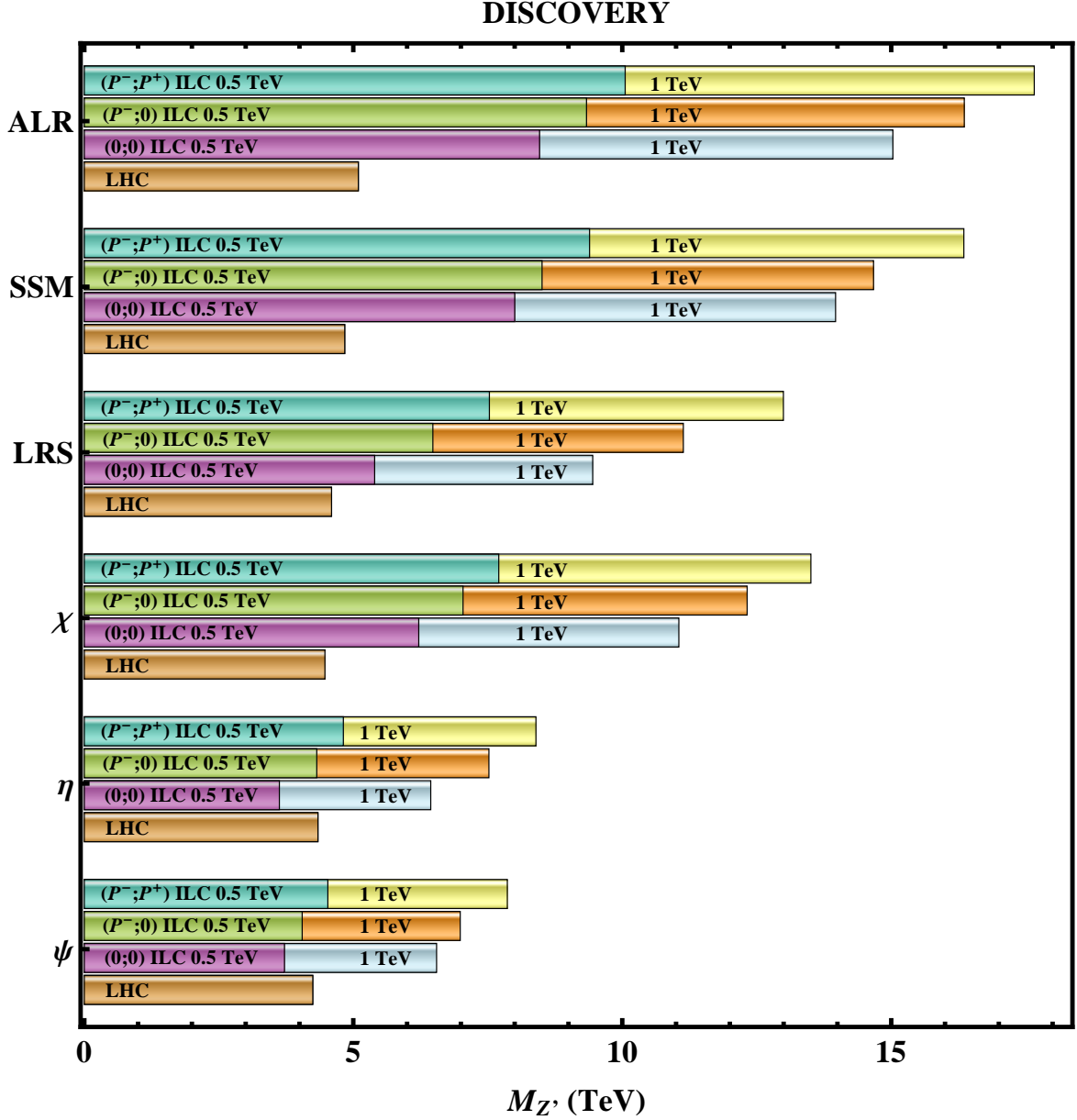}}
\vspace*{1cm} \caption{\label{discovery} Discovery reaches on
$Z^\prime$  models obtained from combined analysis of the
unpolarized and polarized processes (\ref{proc_ff}) (95\% C.L.) at
the ILC with $\sqrt{s}=0.5$~TeV (1 TeV) and $\Lumint=500$
fb$^{-1}$ (1000 fb$^{-1}$), compared to the results expected from
Drell-Yan processes at the LHC at the 5-$\sigma$ level
\cite{Osland:2009tn}. Three options of polarization are considered
at the ILC: unpolarized beams, $P^-= P^+=0$; polarized
electron beam, $\vert P^-\vert=0.8$; both
beams polarized, $\vert P^-\vert=0.8$ and $\vert
P^+\vert=0.6$.}
\end{figure}

Notice that not only the different beam longitudinal
polarizations, but eventually also the various processes `$f$' in
Eq.~(\ref{proc_ff}) are combined in the definition
(\ref{chisquare}). Here, one assumes to have produced a set of
`data', for example by using the dynamics specified by a given
$Z^\prime$ model, and $\delta {\cal O}$
in the denominator denotes the corresponding `experimental'
uncertainty on $\cal O$, combining statistical and,
if possible, systematical ones. According to
the previous considerations, the $\chi^2$, besides the number
of degrees of freedom,
is basically a function of the chosen
$Z^\prime$ model parameters.
In particular, if the coupling constants are fixed at specific
values, it will depend solely on the $Z^\prime$ mass, and we
vary this parameter. The discovery sensitivity to the $Z^\prime$
under consideration can in this
case be identified as the limiting value of $M_{Z^\prime}$ for
which the value of $\chi^2(M_{Z^\prime})$
has the probability needed for {\it exclusion} of the SM at a
desired confidence level (in what follows, we
shall impose 95\% C.L.).
In the cases where $\cos\beta$- or $\alpha_{\rm LR}$-dependent
couplings are considered, SM {\it exclusion} regions can be
defined analogously.

\par
To derive the expected `discovery' limits on $Z^\prime$ models at the
ILC, for the `annihilation' channels in Eq.~(\ref{proc_ff}), with
$f\ne e,t$, we restrict ourselves to combining in
Eq.~(\ref{chisquare}) the $(P^-,P^+)=(\vert P^-\vert,-\vert P^+\vert)$
and $(-\vert P^-\vert,\vert P^+\vert)$ beam polarization
configurations, that are the predominant ones. For the Bhabha process,
$f=e$, we combine in (\ref{chisquare}) the cross sections with all
four possible polarization configurations, i.e.,
$(P^-,P^+)=(\vert P^-\vert,-\vert P^+\vert)$,
$(-\vert P^-\vert,\vert P^+\vert$), $(\vert P^-\vert,\vert P^+\vert)$,
$(-\vert P^-\vert,-\vert P^+\vert)$. Numerically,
following the ILC Design Report \cite{:2007sg}, we take for the
electron beam $\vert P^-\vert=0.8$. For the positron beam, $\vert
P^+\vert=0.3$ is discussed as possibly available `free of charge'
already in the ILC initial running conditions. However, such a small
positron polarization will turn out not to affect our evaluated
discovery and identification reaches on $Z^\prime$s considerably. We
shall therefore present numerical results for two cases, unpolarized
positron beam $\vert P^+\vert=0$, and $\vert P^+\vert=0.6$
representing the `ultimate' upgrade.

\par
Regarding the ILC energy and the time-integrated luminosity
(which, for simplicity, we assume to be equally
distributed
among the different polarization configurations defined above),
still according to Ref.~\cite{:2007sg}, we will give explicit
numerical results for c.m. energy $\sqrt s=0.5$ TeV with
time-integrated luminosity ${\cal L}_{\rm int}= 500$
$\text{fb}^{-1}$, and for
the `ultimate' upgrade values $\sqrt s=1.0$ TeV with
${\cal L}_{\rm int}=1000$ $\text{fb}^{-1}$. The assumed
final state identification efficiencies governing, together with the
luminosity, the expected statistical uncertainties, are: 100\%
for $e^+e^-$ pairs; 95\% for $l^+l^-$ events
($l=\mu,\tau$); 35\% and 60\% for $c {\bar c}$ and
$b {\bar b}$, respectively \cite{:2007sg,Djouadi:2007ik}.

\par
As for the major systematic uncertainties, they originate
from errors on beam polarizations, on the time-integrated
luminosity, and the final-state reconstruction and energy
efficiencies. For the longitudinal polarizations, we
adopt the values $\delta P^-/P^-=\delta P^+/P^+=0.25$\%, rather
ambitious, especially as far as $P^+$ is concerned, but strictly
needed for conducting the planned measurements at the permille
level, see, e.g.,
Refs.~ \cite{Riemann:2009wy,Aurand:2009kp,
Boogert:2009ir}.\footnote{For simplicity we here assume equal
precisions on beam polarizations. Clearly,
much larger systematic errors on $P^+$ could partially spoil the
advantages expected from
the availability of also positron beam polarization.}
As regards the other systematic uncertainties mentioned above,
we assume for the combination the (perhaps conservative)
lumpsum value of 0.5\%. The systematic uncertainties are
included using the covariance matrix approach
\cite{eadie,Cuypers:1996it,Babich:2001nc}.

\begin{figure}[tbh] %
\vspace*{0.0cm} \centerline { \hspace*{-2.0cm}
  \includegraphics[width=11.0cm,angle=0]{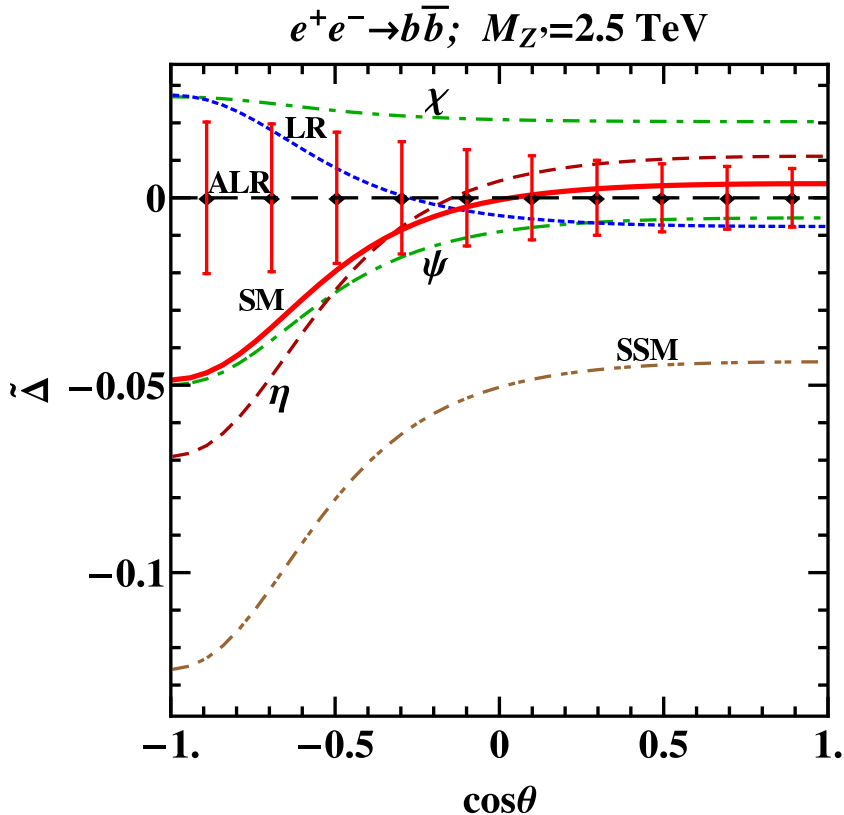}}
\vspace*{0.cm} \caption{\label{deviation} Relative deviation of the
  unpolarized differential cross sections from the ALR-model
  prediction $\tilde\Delta$ for the process $e^+e^- \to b\bar{b}$ as a
  function of $\cos\theta$ for the SM and considered $Z^\prime$ models
  with $M_{Z'}=2.5~\text{TeV}$ at the ILC with
  $\sqrt{s}=0.5~\text{TeV}$ and $\Lumint=500~\text{fb}^{-1}$. The
  error bars are statistical uncertainties at the 1-$\sigma$ level.}
\end{figure}

\par
Concerning the theoretical inputs, for the SM amplitudes we use the
effective Born approximation \cite{Consoli:1989pc}
vertices, with $m_{\rm top}=175~\text{GeV}$ and $m_{\rm H} =
120~\text{GeV}$. The numerically dominant ${\cal O}(\alpha)$ QED
corrections are generated by initial-state radiation, for both Bhabha
scattering and the annihilation processes in (\ref{proc_ff}). They are
accounted for by a structure function approach including both hard and
soft photon emission \cite{nicrosini}, and by a flux factor method
\cite{physicsatlep2}, respectively. Effects of radiative flux return
to the $s$-channel $Z$ exchange are minimized by the cut $\Delta\equiv
E_\gamma/E_{\rm beam}< 1-M_Z^2/s$ on the radiated photon energy, with
$\Delta= 0.9$.  In this way, only interactions that occur close to the
nominal collider energy are included in the analysis and, accordingly,
the sensitivity to the manifestations of the searched-for nonstandard
physics can be optimized. By numerical studies based on the ZFITTER
code \cite{Bardin:1999yd}, other QED effects such as final-state and
initial-final state emission are found, in the
processes $e^+e^-\to l^+l^-$ ($l=\mu,\tau$) and $e^+e^-\to q\bar{q}$
($q=c,b$), to be numerically unimportant for the chosen kinematical
cuts. Finally, correlations between the different polarized cross
sections (but not between the individual angular bins) are taken into
account in the derivation of the numerical results, that we present in
Fig.~\ref{discovery}.  The figure includes a comparison with the
discovery potential of the LHC with luminosity 100~fb$^{-1}$, from the
Drell-Yan processes $pp\to l^+l^- +X$ ($l=e,\,\mu$) (at the 5-$\sigma$
level).  These values provide a representative overview of the
sensitivities of the reach in $M_{Z^\prime}$ on the planned energy and
luminosity, as well as on beam polarization.

\section{Distinction of $Z'$ models}
\label{sect:distinction}

Basically, in the previous subsection we have assessed the extent
to which $Z^\prime$ models can give values of $e^+e^-$
differential cross sections that can {\it exclude} the SM hypothesis
to a prescribed C.L. Such `discovery reaches' are represented by
upper limits on $Z^\prime$ masses, for which the observable
deviations between the corresponding $Z^\prime$ models and SM
predictions are sufficiently large compared to the foreseeable
experimental uncertainties on the cross sections at the ILC.

\par
However, since different models can give rise to
similar deviations, we would like to determine the
ILC potential of identifying, among the various competing
possibilities, the source of a deviation, should it be
effectively observed. These ID-limits should obviously
be expected to lie below the corresponding ILC discovery
reaches and, for an approximate but relatively simple assessment,
we adapt the naive $\chi^2$-like procedure applied in the previous
subsection.

\par
To this purpose, we start by defining a `distance' between pairs of
$Z^\prime$ models, $i$ and $j$ with $i,j$ denoting any of the SSM, SM,
ALR, LRS, $\psi$, $\eta$, $\chi$, but $i\ne j$.
We assume for example model $i$ to be
the `true' model, namely, we consider `data' sets obtained from the
dynamics $i$, with corresponding `experimental' uncertainties,
compatible with the expected `true' experimental data. The assessment
of its distinguishability from a $j$ model, that we call `tested'
model, can be performed by a $\chi^2$ comparison analogous to
(\ref{chisquare}), with the $\chi^2$ defined as:
\begin{equation}
\chi^2({\cal O})_{i,j}=\sum_{f}\sum_{\{P^-,\ P^+\}} \sum_{\rm
bins}\frac{\left[{\cal O}(Z^\prime_i) - {\cal
O}(Z^\prime_j)\right]_{\rm bin}^2}{\left(\delta_i{\cal O}_{\rm
bin}\right)^2}. \label{chisquareprime}
\end{equation}

\par
As an illustration, the angular behavior of the deviations in the
numerator of Eq.~(\ref{chisquareprime}) for the unpolarized
annihilation $e^+e^-\to b\bar b$ is depicted in Fig.~\ref{deviation},
for the case where the `true' model is $i=$ ALR, with
$M_{Z^\prime}=2.5$ TeV for all models, at the ILC with
$\sqrt s=0.5$ TeV and ${\cal L}_{\rm int}=500\ {\rm fb}^{-1}$
(actually, in this figure, $\tilde\Delta$ is the relative deviation,
$\tilde\Delta=\dd\sigma(Z'_{ALR})/\dd\sigma(Z'_j)-1$).

\par
Basically, considering that the ILC will start when the LHC will
already be operating at the design energy and luminosity, as
anticipated previously, we can envisage two cases requiring
somewhat different strategies.

\subsection{$Z'$ mass known}

In the first case we assume that the $Z^\prime$ mass is already
measured at the LHC, but perhaps not `identified' there, and the value
is within the ILC discovery reaches for both models $i$ and $j$. In
this case one should set $M_{Z^\prime_i}=M_{Z^\prime_j}\equiv
M_{Z^\prime}$ in Eq.~(\ref{chisquareprime}) and, accordingly, the
$\chi^2$ becomes a function of only the $Z^\prime_i$ and $Z^\prime_j$
coupling constants. If both the $Z^\prime_i$ and $Z^\prime_j$
couplings are fixed numerically, like in the example of
Fig.~\ref{deviation}, distinguishability can be assessed by varying
$M_{Z^\prime}$, up to the point where the $\chi^2_{ij}$ reaches the
critical value suitable for {\it exclusion} of the `tested' model $j$
by the `true' model $i$ at the desired confidence level.

If the above mentioned couplings are, instead, the $\beta$- or
$\alpha_{\rm LR}$- dependent ones, `confusion' domains between `true'
and `tested' models can analogously be determined by means of
Eq.~(\ref{chisquareprime}) in the model parameter plane
$(\cos\beta,\alpha_{\rm LR})$ for fixed values of
$M_{Z^\prime_i}=M_{Z^\prime_j}\equiv M_{Z^\prime}$. By definition,
in these `confusion' domains, that depend on the actual value assumed
for $M_{Z^\prime}$, the cross sections corresponding to
definite values of $\beta$ and $\alpha_{\rm LR}$ cannot be
distinguished from each other at the desired confidence level.
Correspondingly, the `complementary' regions in the above mentioned
parameter plane can define the `resolution' domain of the `tested'
model by the the `true' model hypothesis, and determine in this way
the identification limit on the latter.

\begin{figure}[htbp]
\centerline{ \hspace*{-0.0cm}
\includegraphics[width=6.4cm,angle=0]{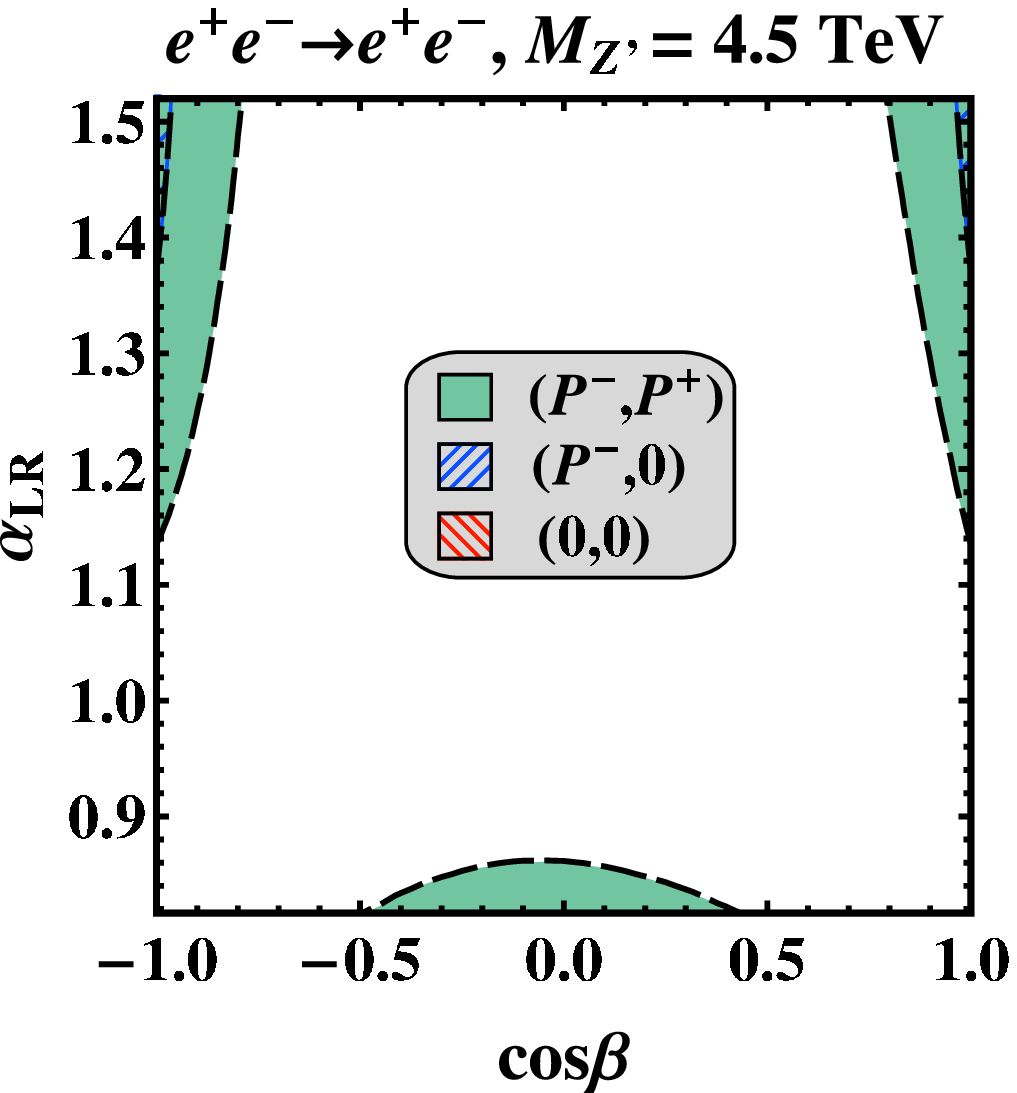}
\hspace*{-0.2cm}
\includegraphics[width=6.4cm,angle=0]{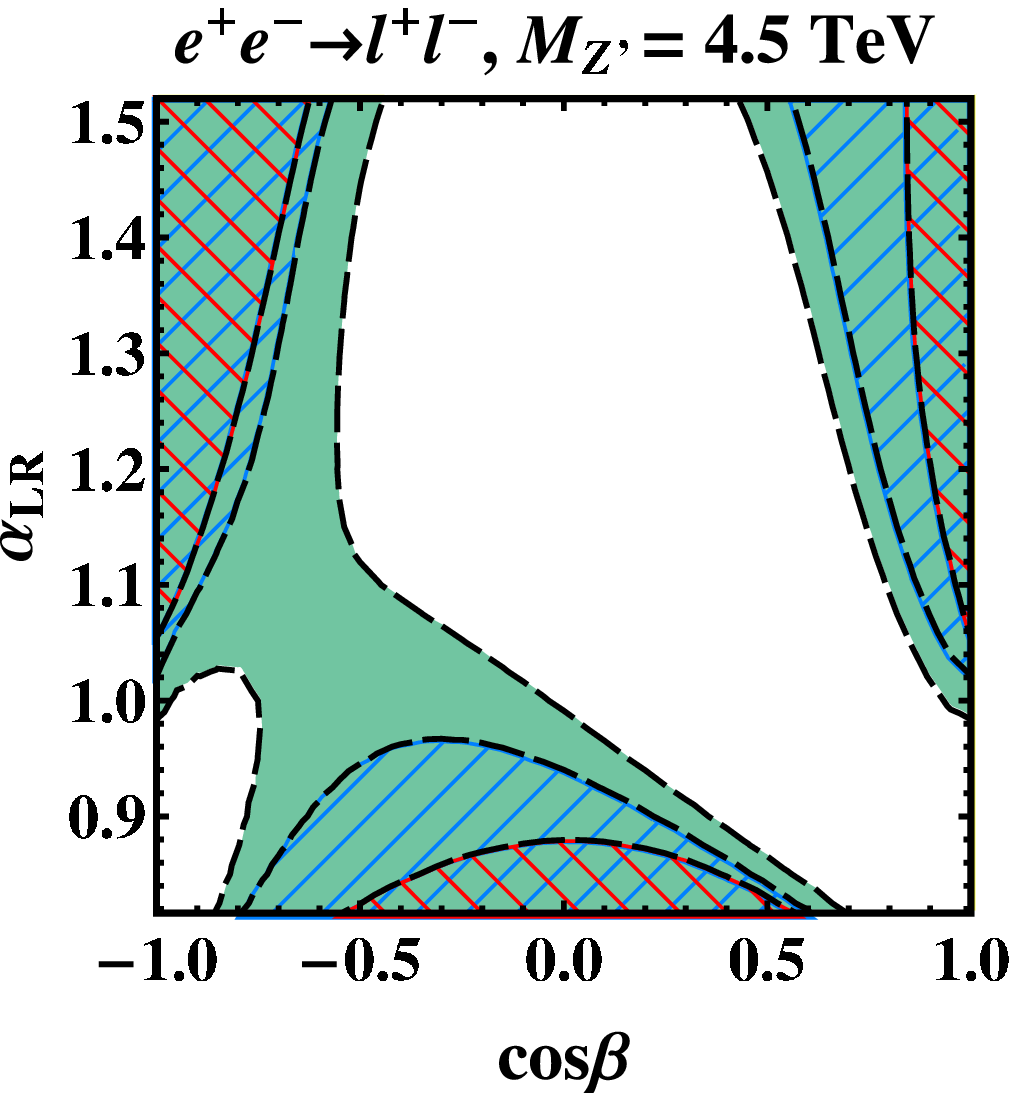}}
\centerline{ \hspace*{-0.2cm}
\includegraphics[width=6.4cm,angle=0]{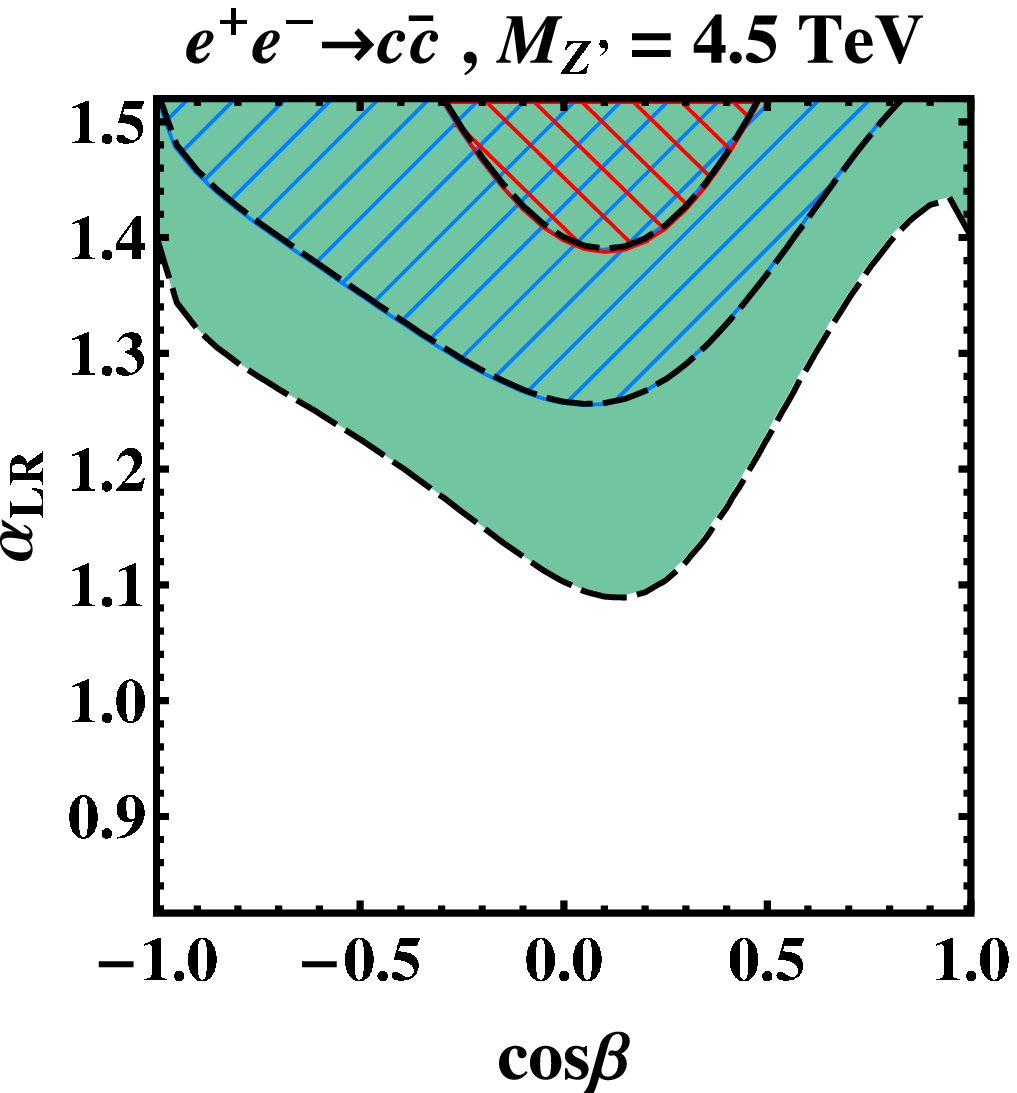}
\hspace*{-0.2cm}
\includegraphics[width=6.4cm,angle=0]{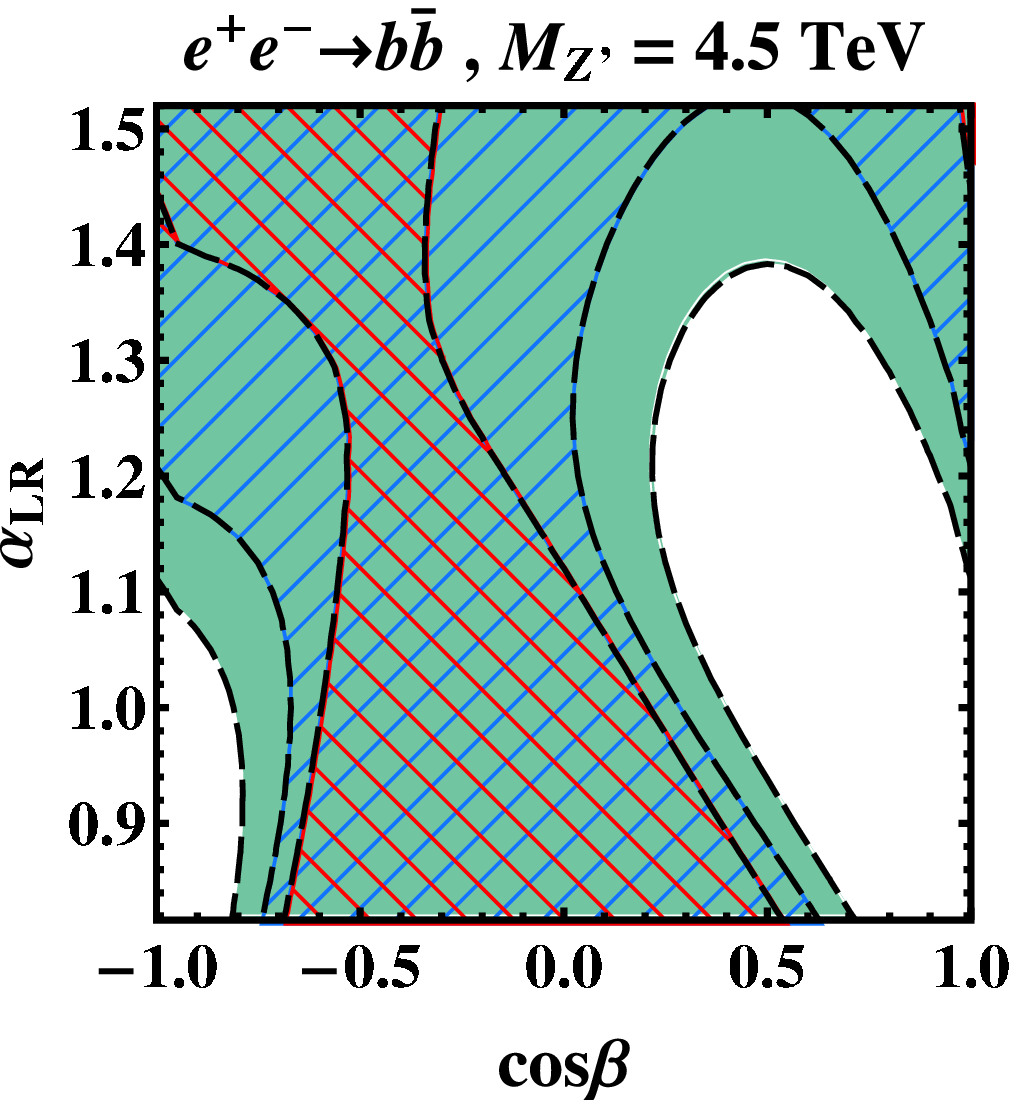}}
\caption{\label{fig4500} Regions of resolution (red hatched) of $E_6$ vs.\
  left-right LR models in the model parameter plane
  ($\cos\beta,\alpha_{\rm LR}$) for $M_{Z'}=4.5~\text{TeV}$ obtained
  with Eq.~(\ref{chisquareprime}) from the processes with different
  final states $e^+e^-\to e^+e^-$, $l^+l^-$ ($l=\mu,\,\tau$)
  $c\bar{c}$, $b\bar{b}$, at $\sqrt{s}=0.5~\text{TeV}$,
  $\Lumint= 500~\text{fb}^{-1}$. The role of polarization is
  demonstrated. Using also a
  polarized electron (and positron) beam, the resolution region
  would enlarge to include also the blue hatched (plus the shaded, unhatched) regions, but
  not the white ones. Those would remain as `confusion' regions.}
\end{figure}

\begin{figure}[htb]
\centerline{ \hspace*{-0.0cm}
\includegraphics[width=6.4cm,angle=0]{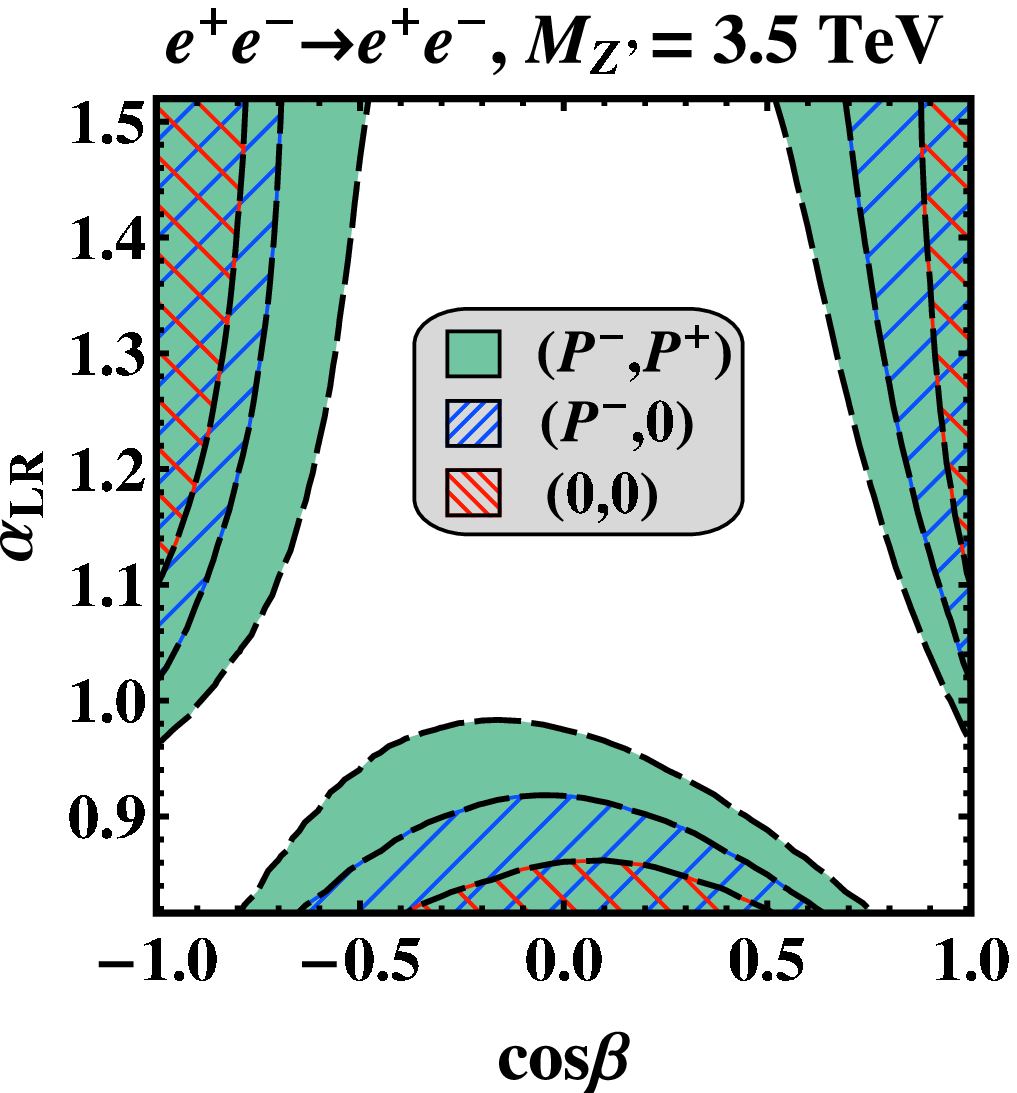}
\hspace*{-0.2cm}
\includegraphics[width=6.6cm,angle=0]{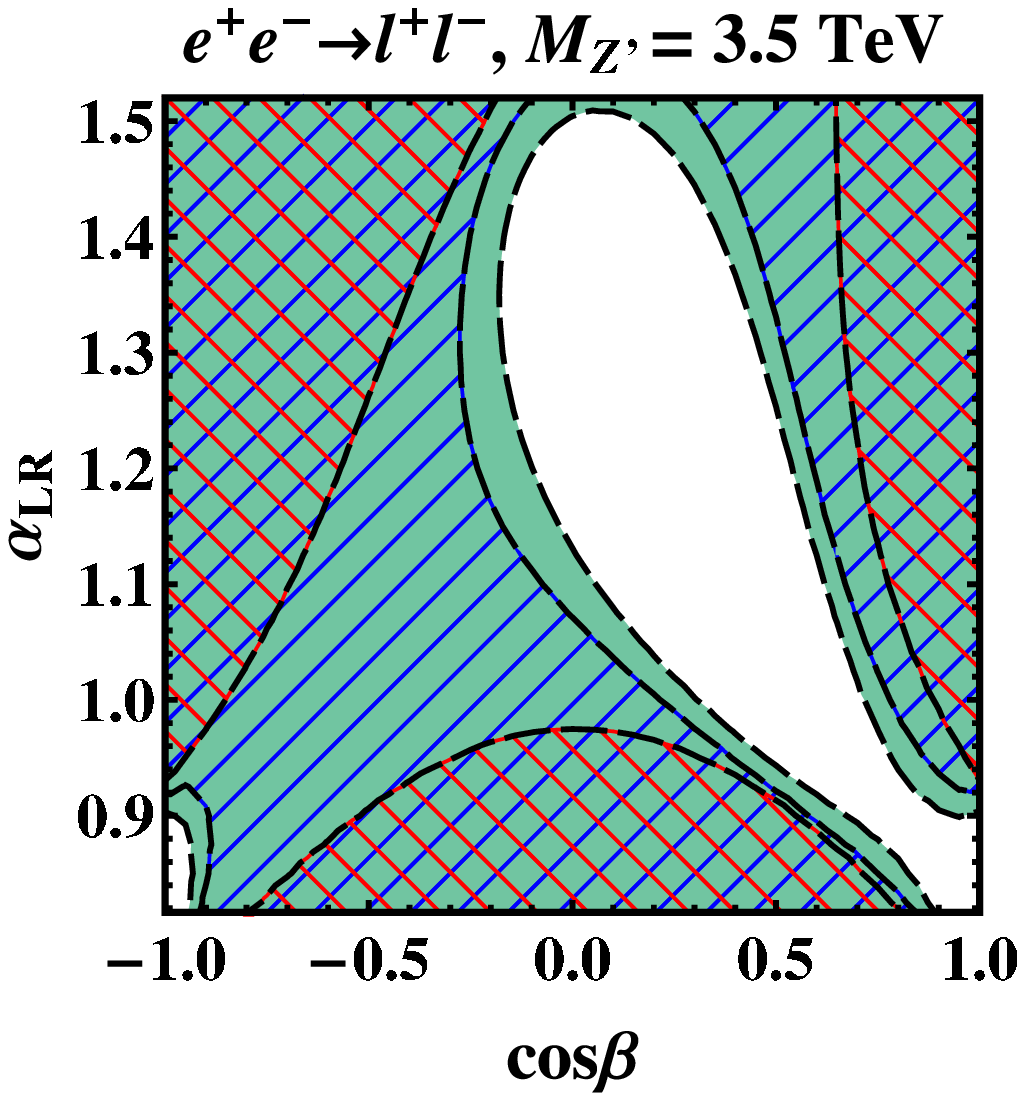}} 
\centerline{ \hspace*{-0.2cm}
\includegraphics[width=6.4cm,angle=0]{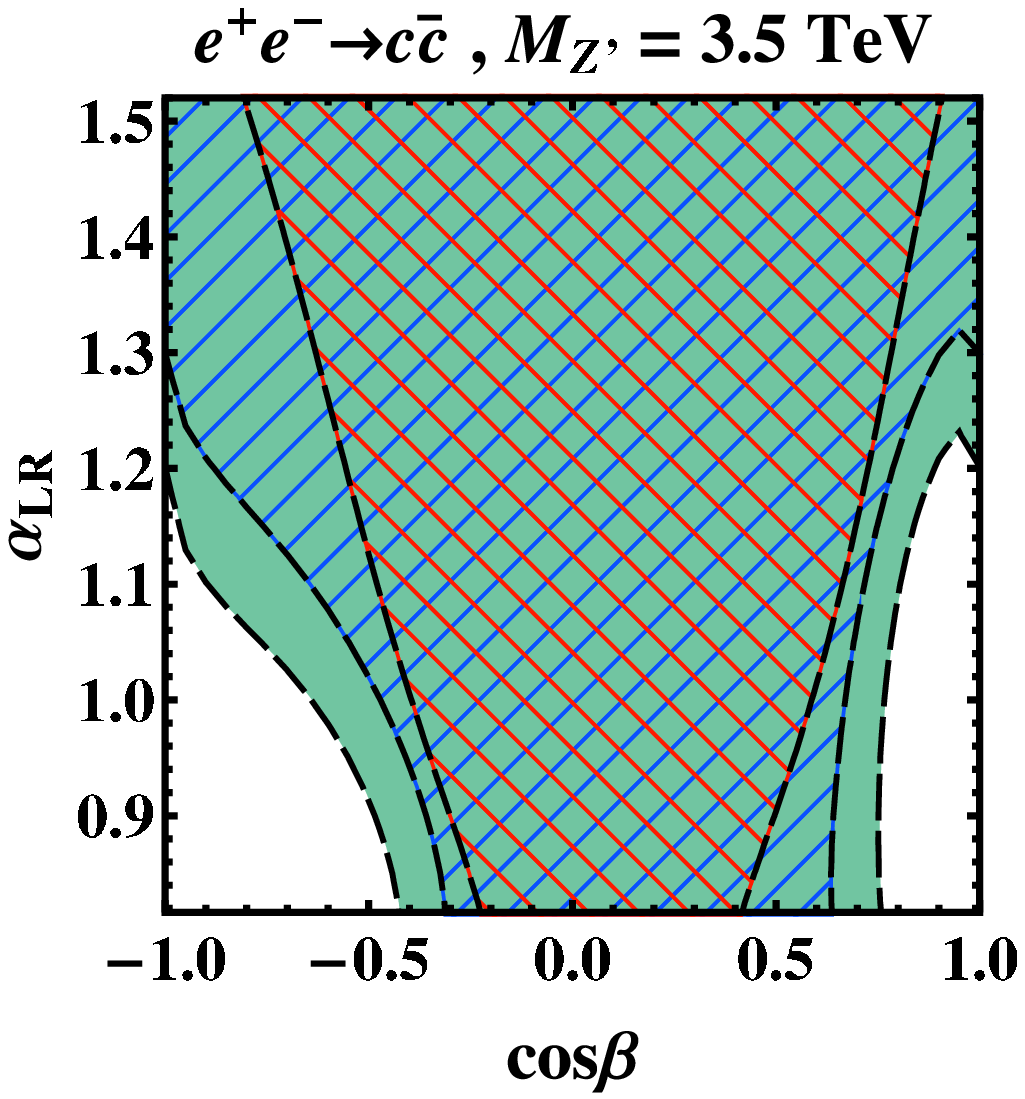}
\hspace*{-0.2cm}
\includegraphics[width=6.4cm,angle=0]{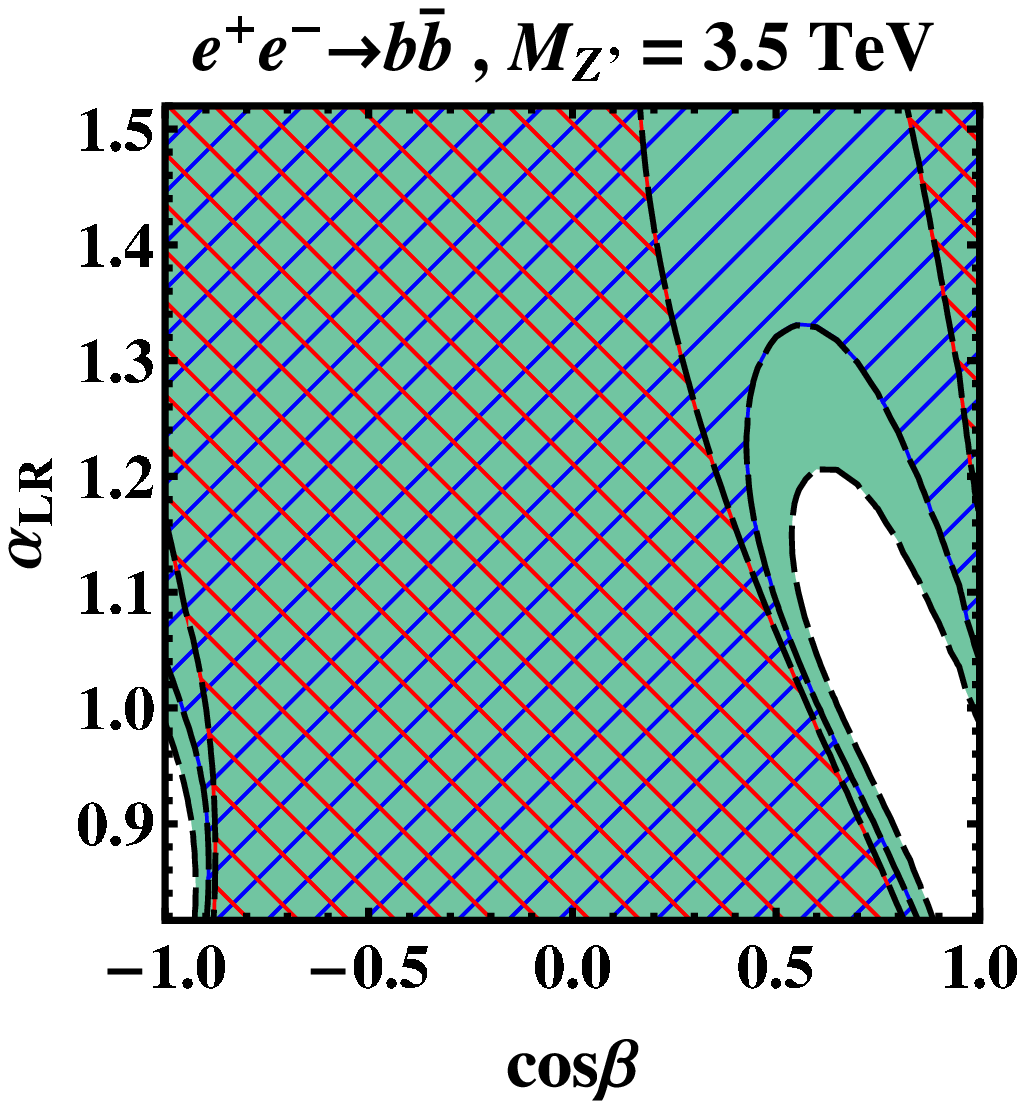}}
\caption{\label{fig3500} Similar to Fig.~\ref{fig4500} but for
$M_{Z'}=3.5~\text{TeV}$.}
\end{figure}

\begin{figure}[htb]
\centerline{ \hspace*{-0.0cm}
\includegraphics[width=6.4cm,angle=0]{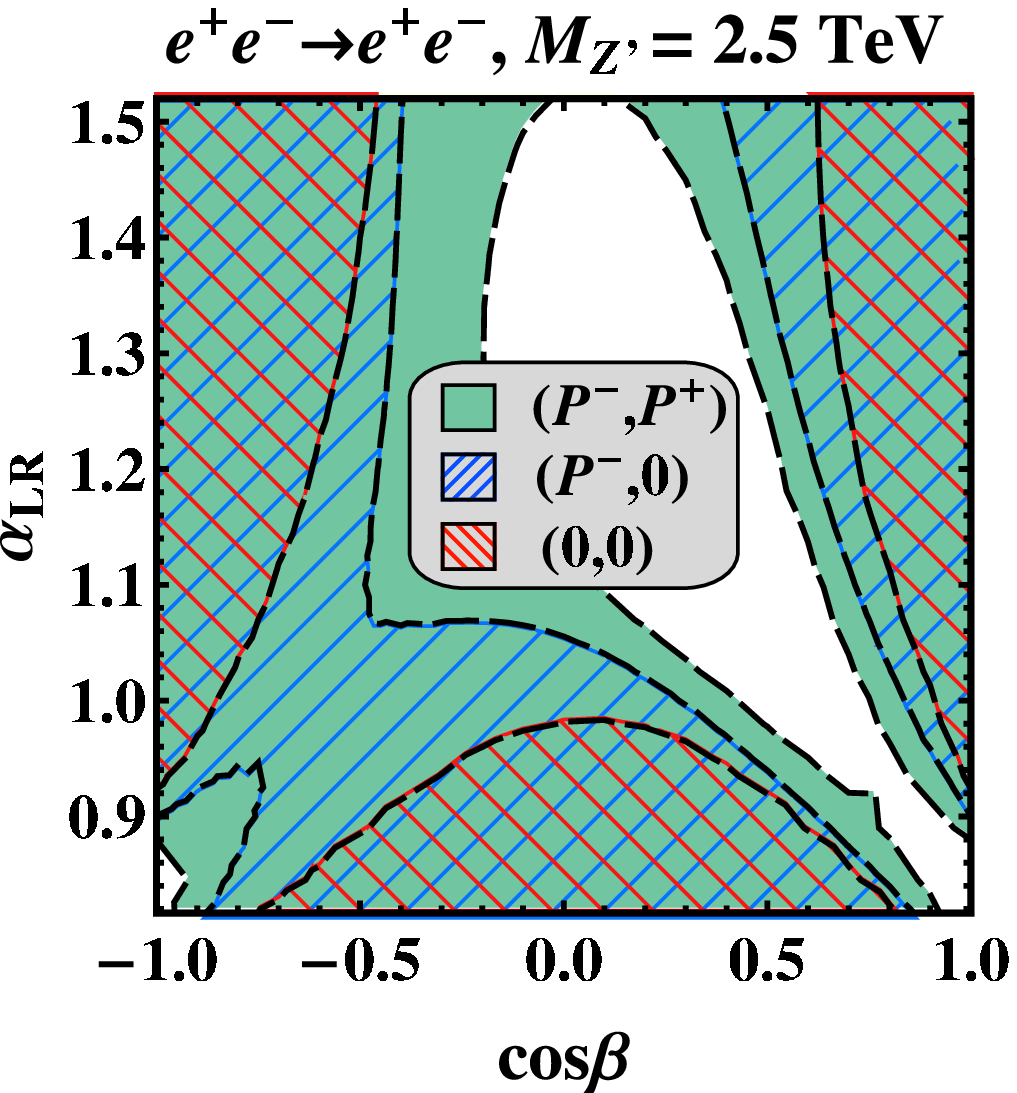}
\hspace*{-0.2cm}
\includegraphics[width=6.4cm,angle=0]{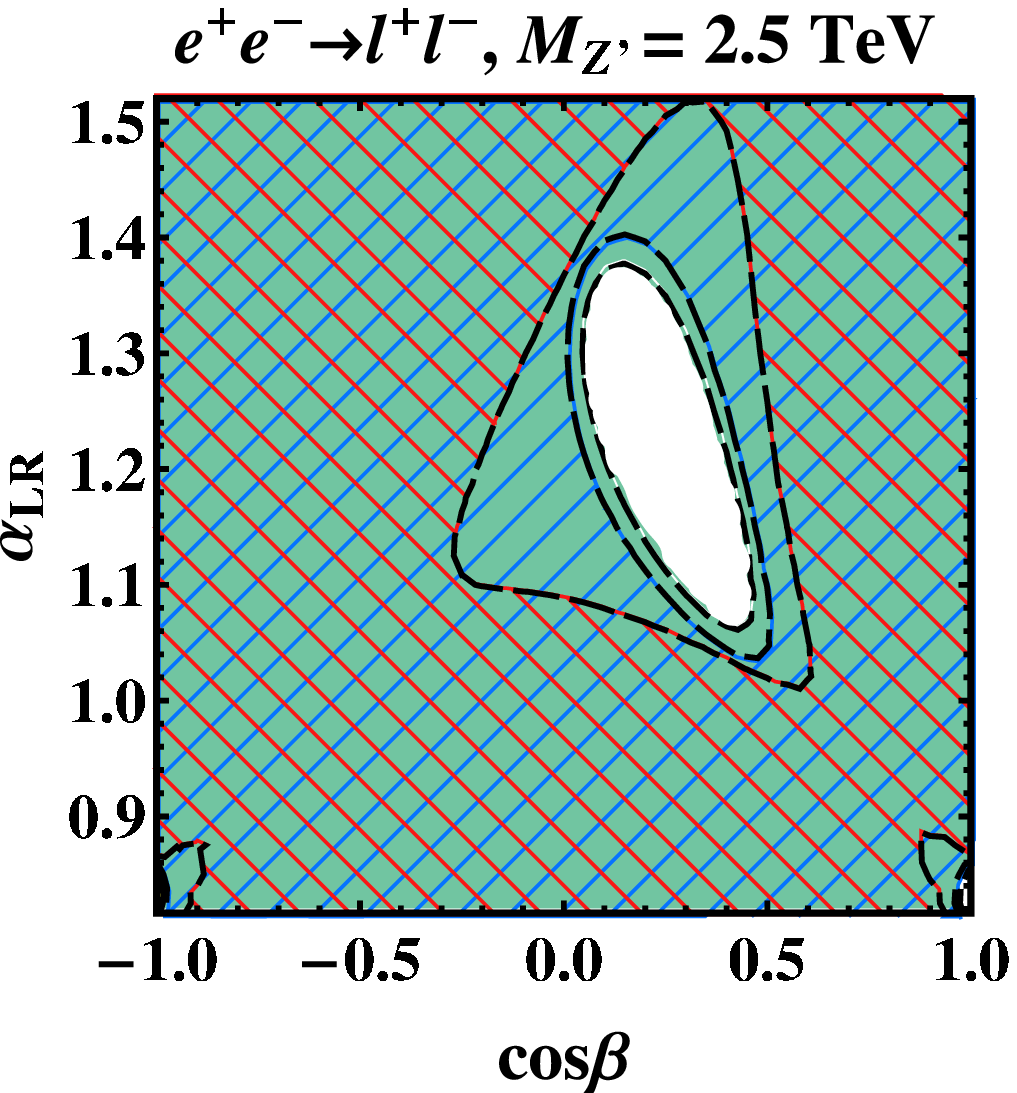}}
\centerline{ \hspace*{-0.2cm}
\includegraphics[width=6.4cm,angle=0]{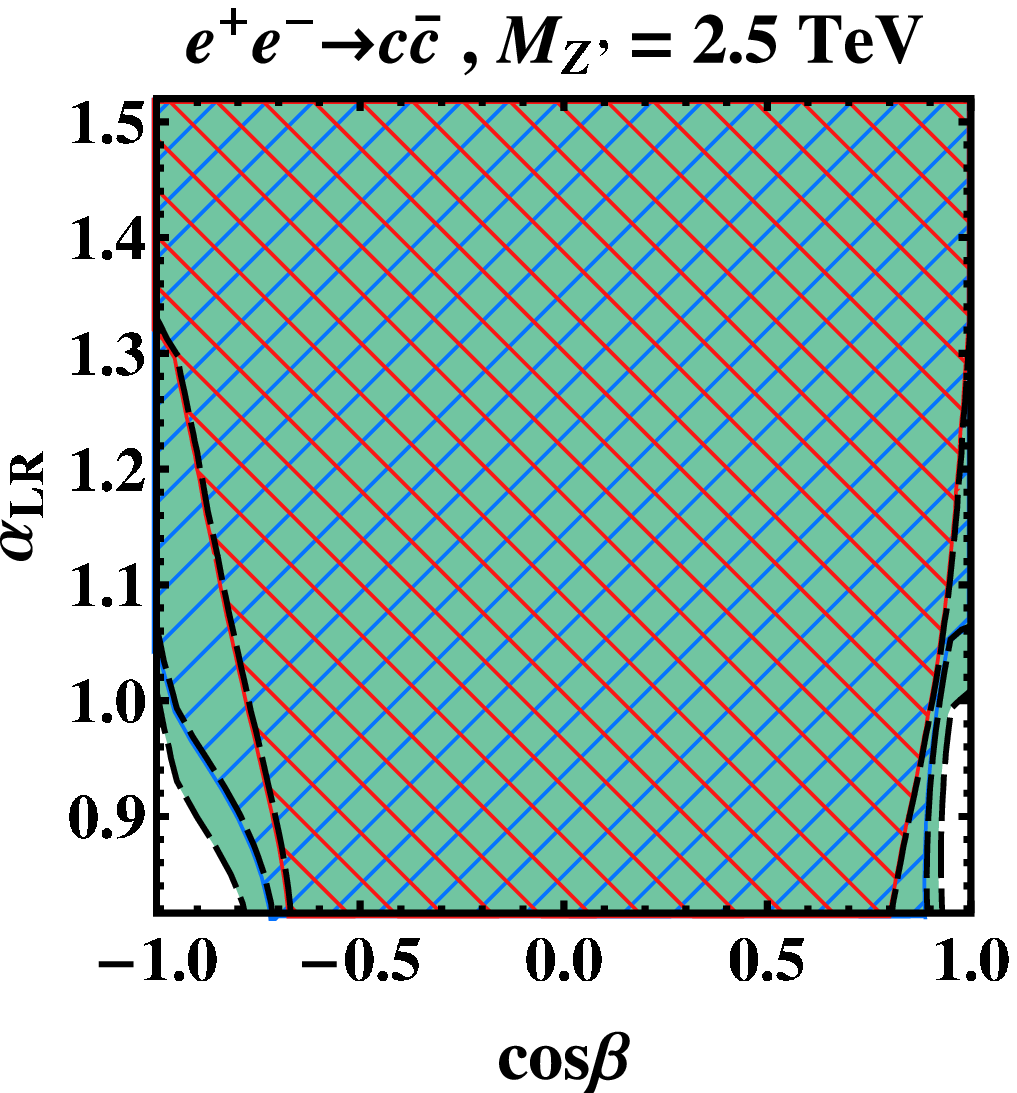}
\hspace*{-0.2cm}
\includegraphics[width=6.4cm,angle=0]{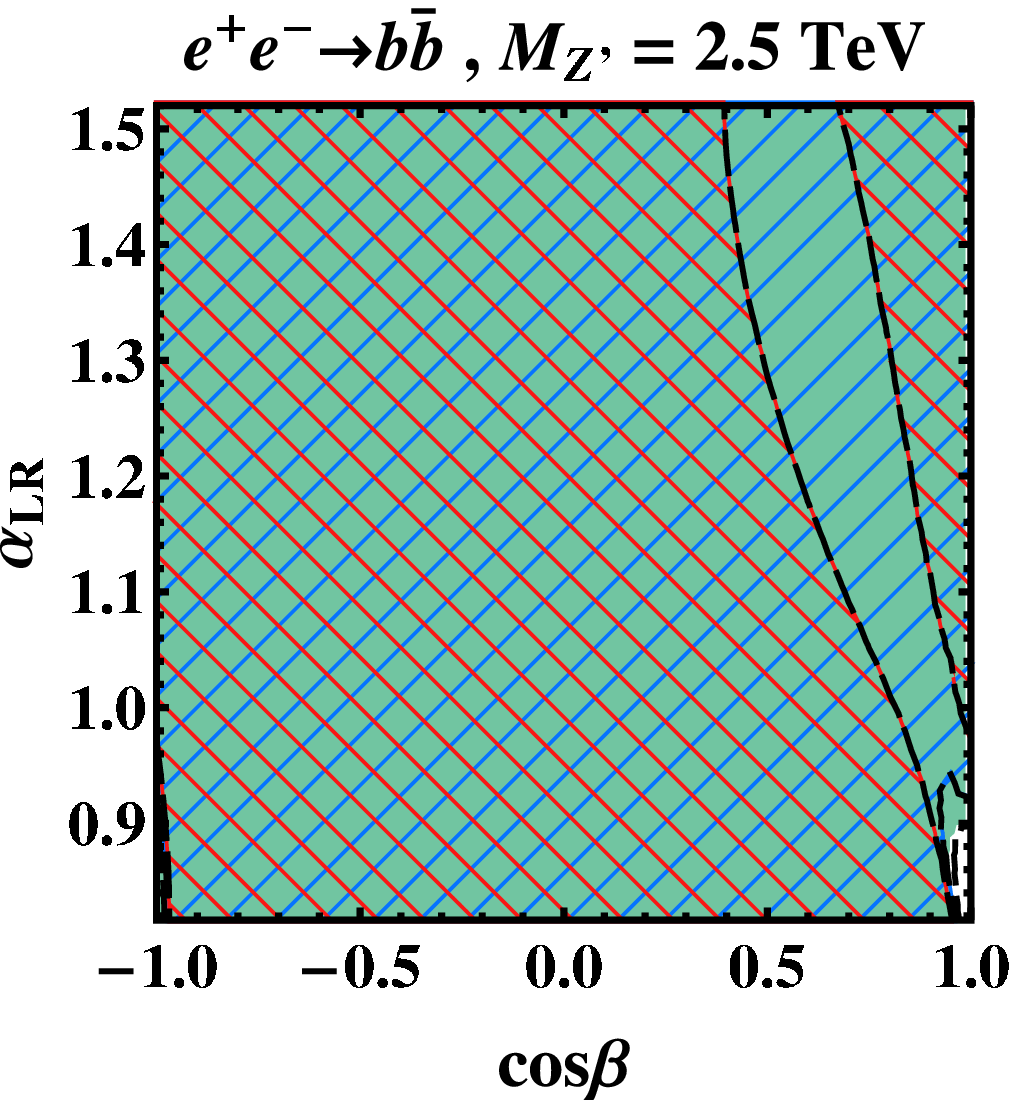}}
\caption{\label{fig2500} Similar to Fig.~\ref{fig4500} but for
$M_{Z'}=2.5~\text{TeV}$.}
\end{figure}

\par
\begin{figure}[htb]
\centerline{
\includegraphics[width=6.4cm,angle=0]{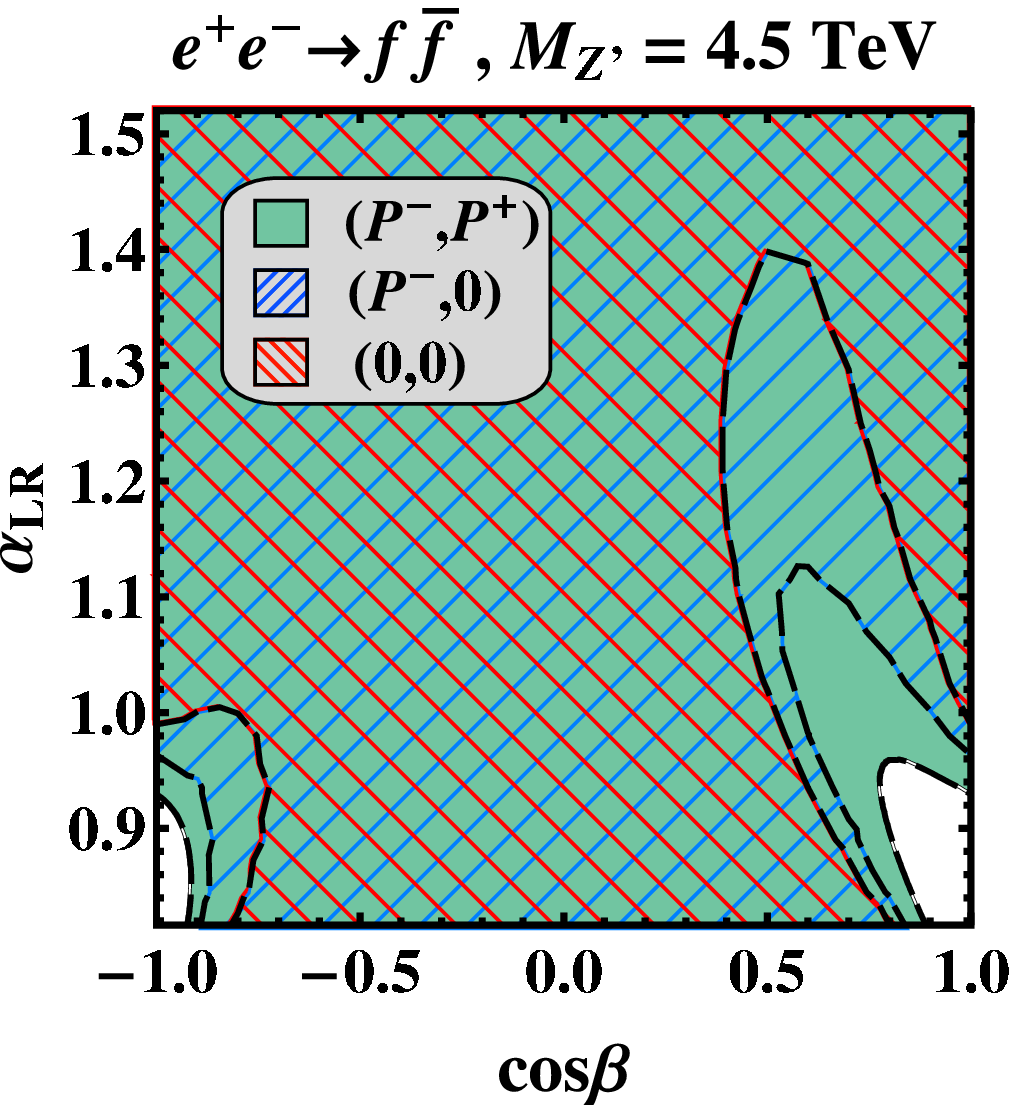}
\includegraphics[width=6.65cm,angle=0]{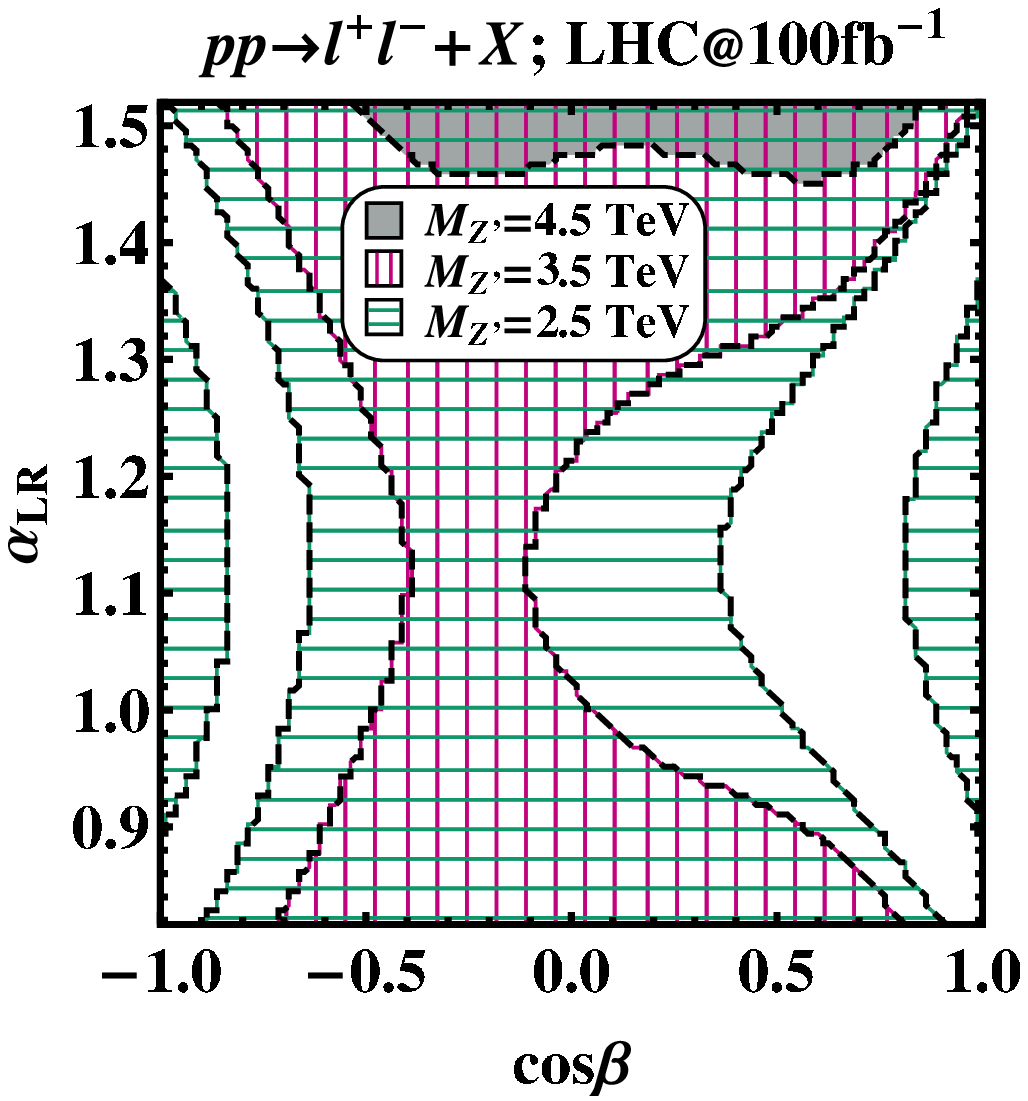}}
\caption{\label{fig-total} Left: Resolution domains
at $M_{Z^\prime}=4.5$, exploiting all final states.
Right: Corresponding resolution domains for Drell-Yan production
at the LHC, for the three values of $M_{Z^\prime}$. }
\end{figure}

As an illustrative example of application of the ID-criteria exposed
above, we evaluate the `resolution' regions between $E_6$ (`true') and
left-right LR (`tested') models in the plane
($\cos\beta,\alpha_{\rm LR}$) for different values of
$M_{Z'}$. Figures~\ref{fig4500}-\ref{fig-total} show the regions of
`resolution' obtained from the processes $e^+e^-\to e^+e^-$, $l^+l^-$
($l=\mu,\,\tau$), $c\bar{c}$ and $b\bar{b}$, for
$M_{Z'}=$4.5 TeV, 3.5 TeV and 2.5 TeV at $\sqrt{s}=0.5$ TeV and
$\Lumint=$ 500 fb$^{-1}$, and for different values of beam
polarization. Notice that, in these figures, the horizontal axis
includes also the values of $\beta$ specific of the $\chi$, $\psi$
and $\eta$ models, while the vertical axis includes the value of
$\alpha_{\rm LR}$ representative of the LRS model.\footnote{Actually,
we should recall that the point
$(\cos\beta,\alpha_{\rm LR})=(1,\sqrt{2/3})$ corresponds to the
$\chi$ model which, indeed, exists in both classes of $Z^\prime$
models. Therefore, since one would be testing the $\chi$ model
against itself, that point must not be considered in an analysis
based on Eq.~(\ref{chisquareprime}).}

\par
Figures~\ref{fig4500}--\ref{fig2500} clearly demonstrate the
complementary roles of the processes with different final states, in
particular, that the process $e^+e^-\to b\bar{b}$ can potentially be
the most efficient one in distinguishing $E_6$ and left-right models
from each other (it provides the largest resolution
domains). Conversely, the purely leptonic processes,
$f=e,\,\mu,\,\tau$ in (\ref{proc_ff}), turn out to determine much less
extended `resolution' areas, in particular they cannot discriminate
the LR model from the $E_6$ models at $(\cos\beta,\alpha_{\rm
  LR})=(\pm 1,\sqrt{2/3})$ and $(1/4, \sqrt{3/2})$ (for any
$M_{Z'}>\sqrt{s}$).

Also, as can be seen from these figures, the leptonic processes
are found to provide `confusion' domains (white) located in the `central'
part of the plane $(\cos\beta,\alpha_{\rm LR})$, around
$(1/4, \sqrt{3/2})$, whereas the processes into $q\bar q$ final states
exhibit the opposite feature. Therefore, as shown in
Fig.~\ref{fig-total}, the combination of
all processes $f$ is expected to dramatically reduce the `confusion'
area in the above mentioned plane and to determine the largest
possible domain in which the considered $Z^\prime$ models can be
mutually distinguished from one another. The substantial role of
electron polarization and, to a somewhat lesser
extent, of positron polarization in shrinking `confusion' domains,
leading to enlarged model `resolution' domains, can also
be seen in these panels. Combining all processes, and with both beams
polarized, a `confusion' turns out to persist only in the minute corners
shown in Fig.~\ref{fig-total}, for $M_{Z^\prime}=4.5~\text{GeV}$,
and nothing at the lower masses.

\begin{figure}[htb]
\centerline{ \hspace*{-0.5cm}
\includegraphics[width=7.6cm,angle=0]{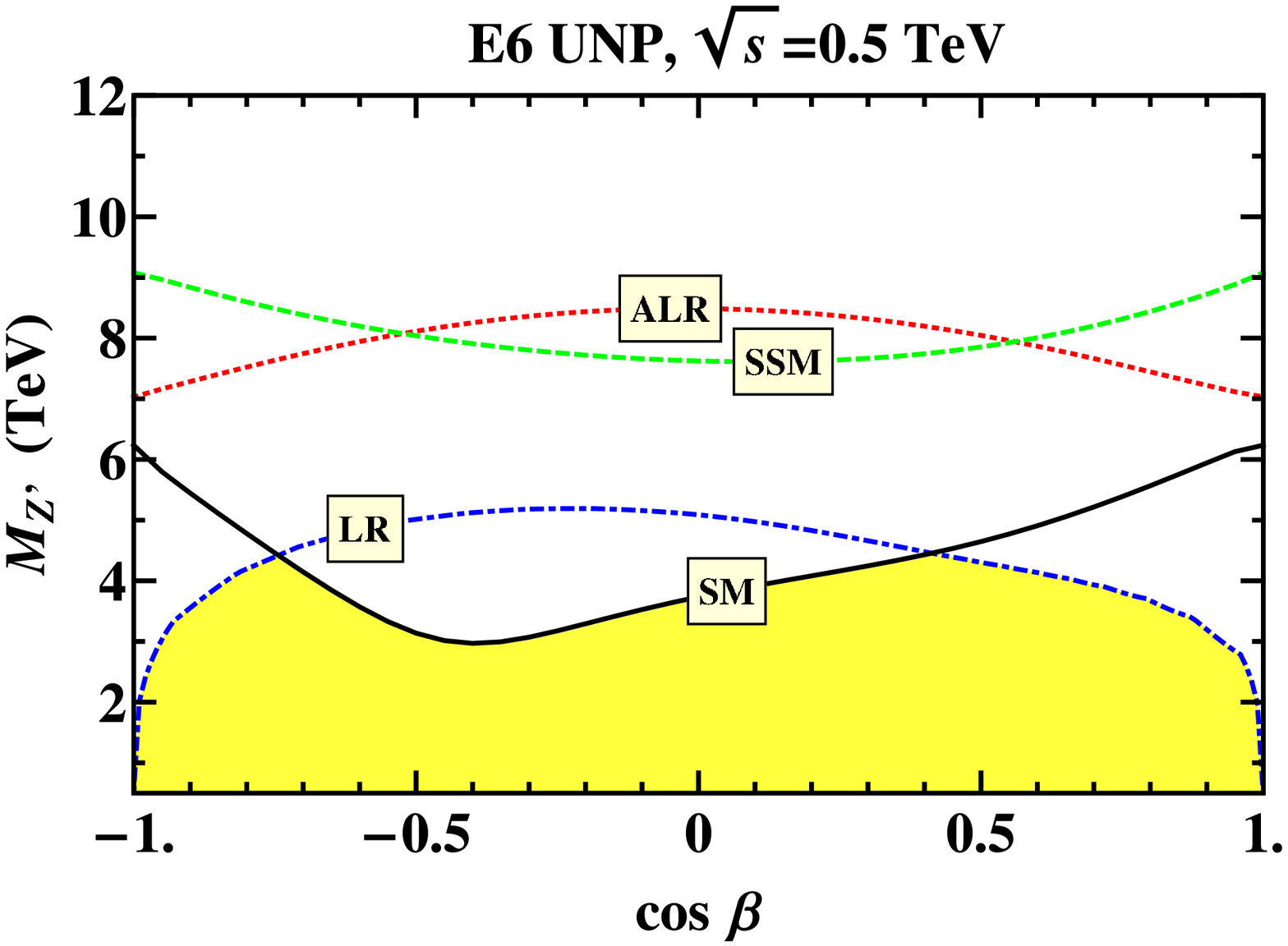}
\hspace*{0.2cm}
\includegraphics[width=7.6cm,angle=0]{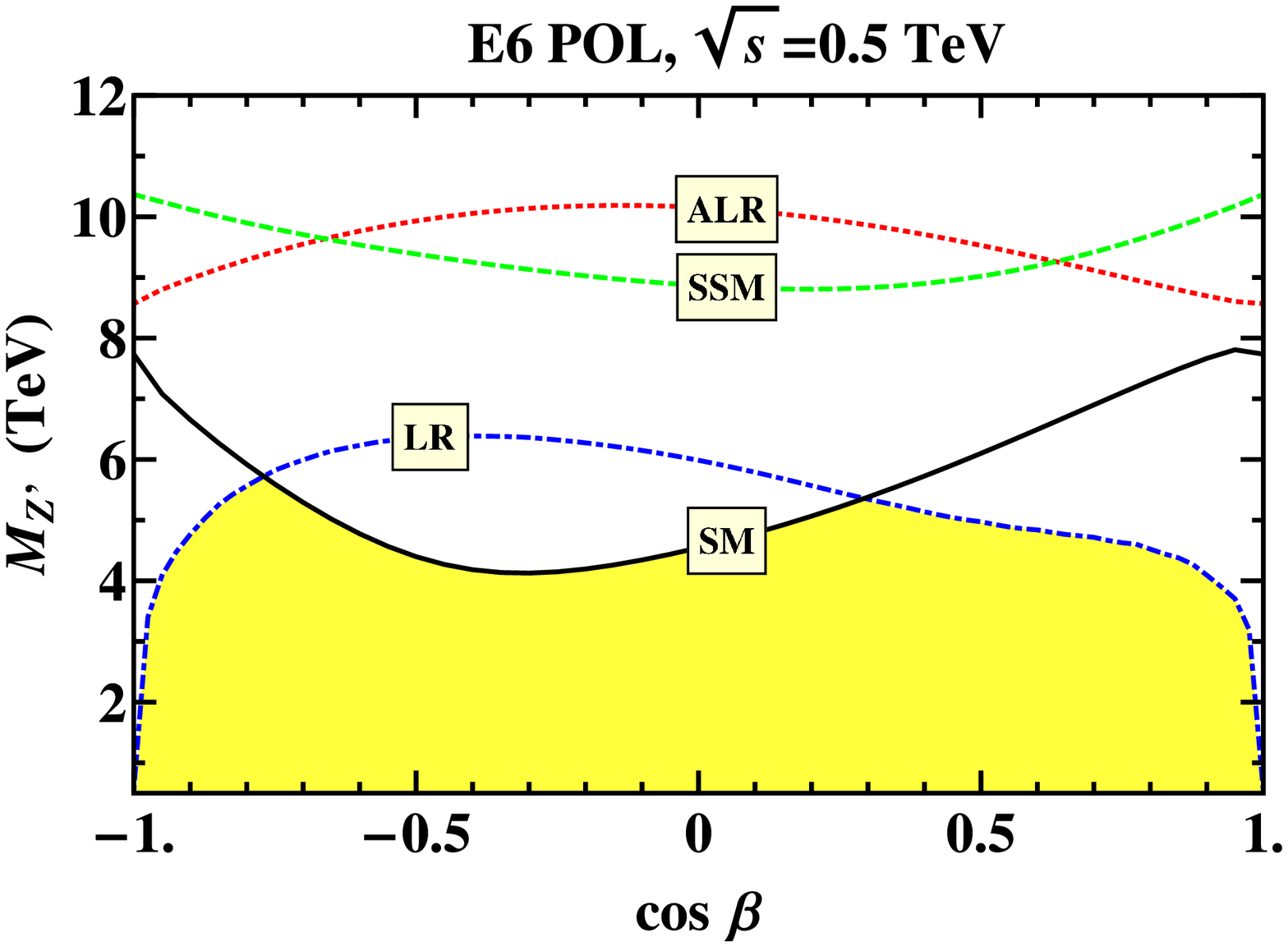}}
\caption{\label{ID-E6}{ Left panel:} E6 identification reaches on
  $M_{Z'}$ at 95\% C.L. obtained from combination of all unpolarized
  processes $e^+e^- \to f\bar{f}$ at $\sqrt{s}$=0.5 TeV and
  $\Lumint$=500 fb$^{-1}$. The $E_6$ model is assumed to be `true'
  while the others (SSM, LR, ALR, SSM and SM) are taken as tested
  models. The identification range is indicated as the shaded (yellow)
  area. {Right panel:} Similar, but for the polarized processes.}
\end{figure}

\begin{figure}[htb]
\centerline{ \hspace*{-0.5cm}
\includegraphics[width=7.6cm,angle=0]{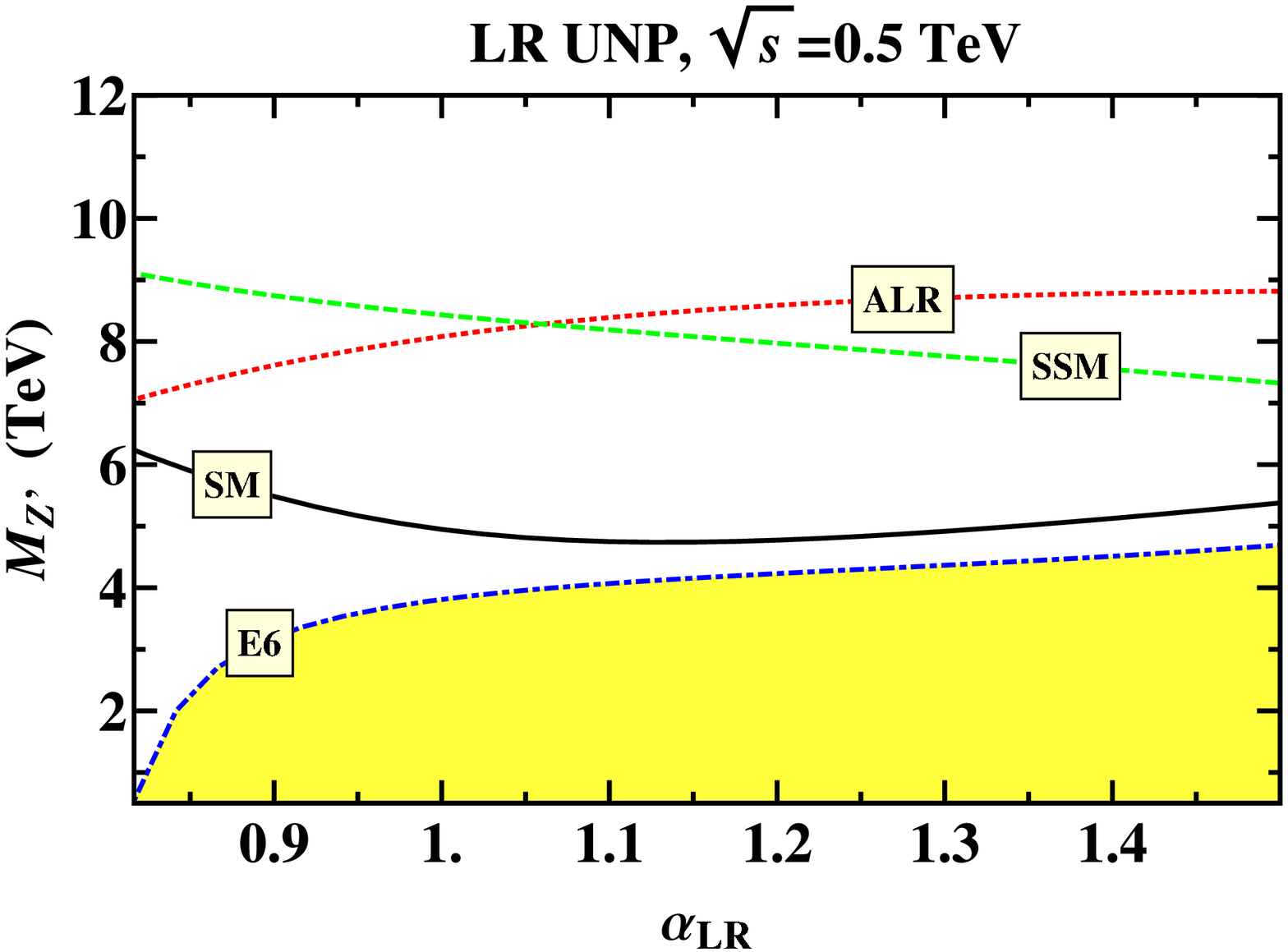}
\hspace*{0.2cm}
\includegraphics[width=7.6cm,angle=0]{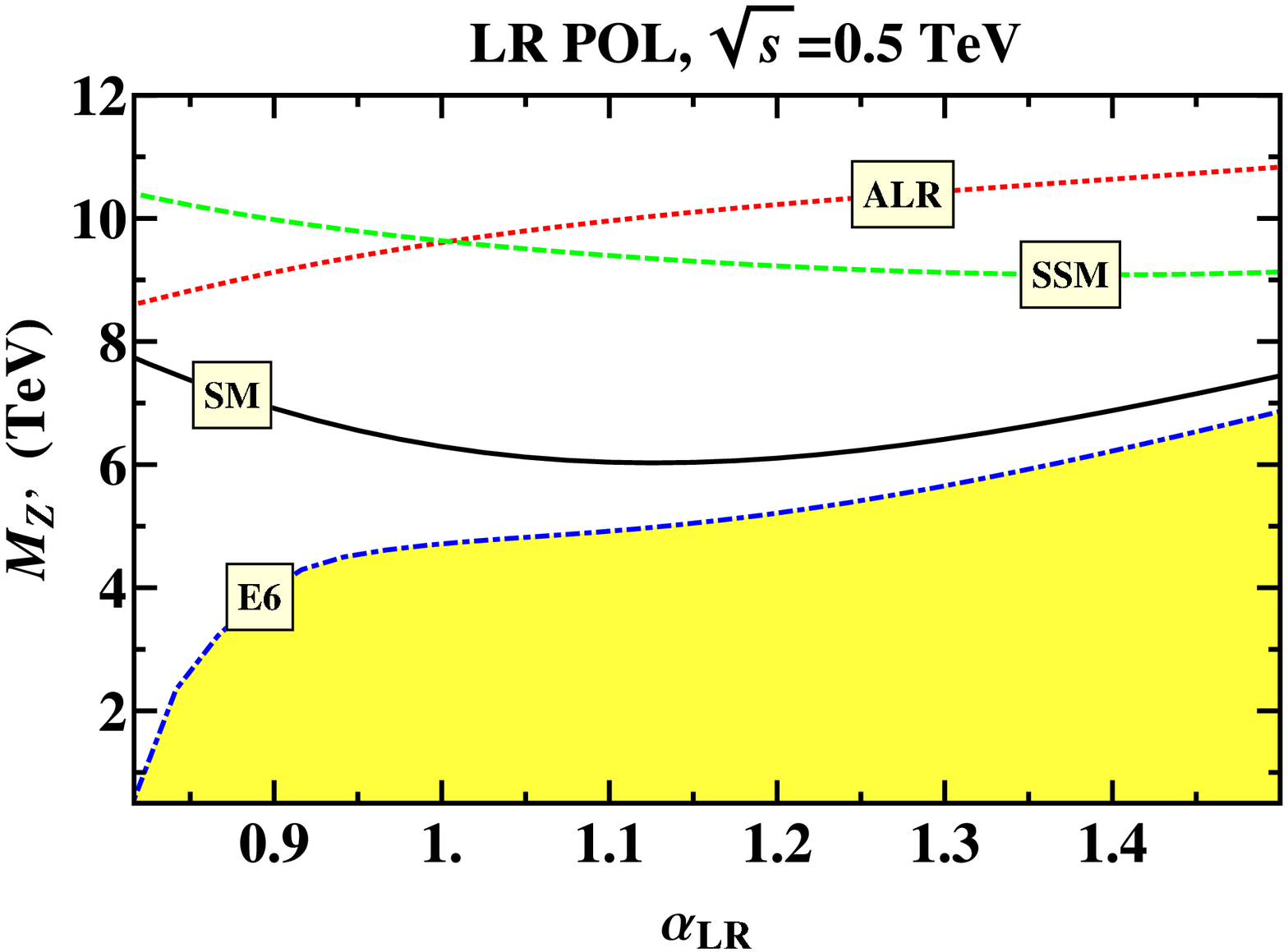}}
\caption{\label{ID-LR} Similar to Fig.~\ref{ID-E6}, for LR
models.}
\end{figure}

It is interesting to compare these resolution regions with the
corresponding ones resulting from the assumed $Z^\prime$ discovery
in the Drell-Yan process at the LHC, shown in the lower-right
panel of Fig.~\ref{fig-total}. This figure shows that, at the LHC, for the
discovery of a 4.5~TeV $Z^\prime$ the corresponding resolution
region is found to cover only a narrow strip,
$1.4\lsim\alpha_\text{LR}$ and $-0.5\lsim\cos\beta\lsim 0.8$. Even
for $M_{Z^\prime}=2.5~\text{TeV}$, at the LHC, there are only modest
hyperbola-like strips at $|\cos\beta|\gsim0.5$ where models can
be distinguished.

\begin{figure}[htb]
\centerline{ 
\includegraphics[width=10.5cm,angle=0]{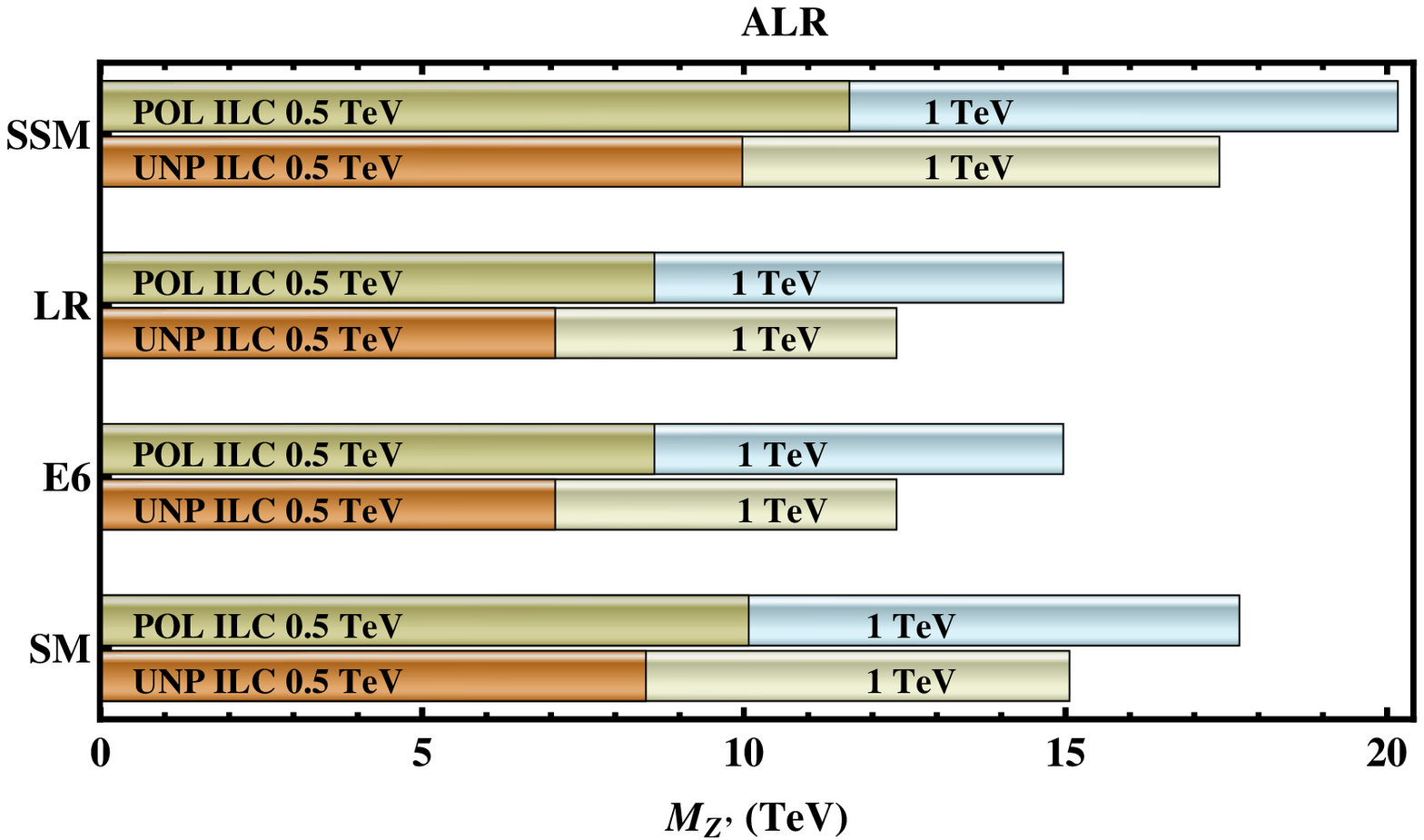}}
\vspace*{0.5cm}
\centerline{
  \includegraphics[width=10.5cm,angle=0]{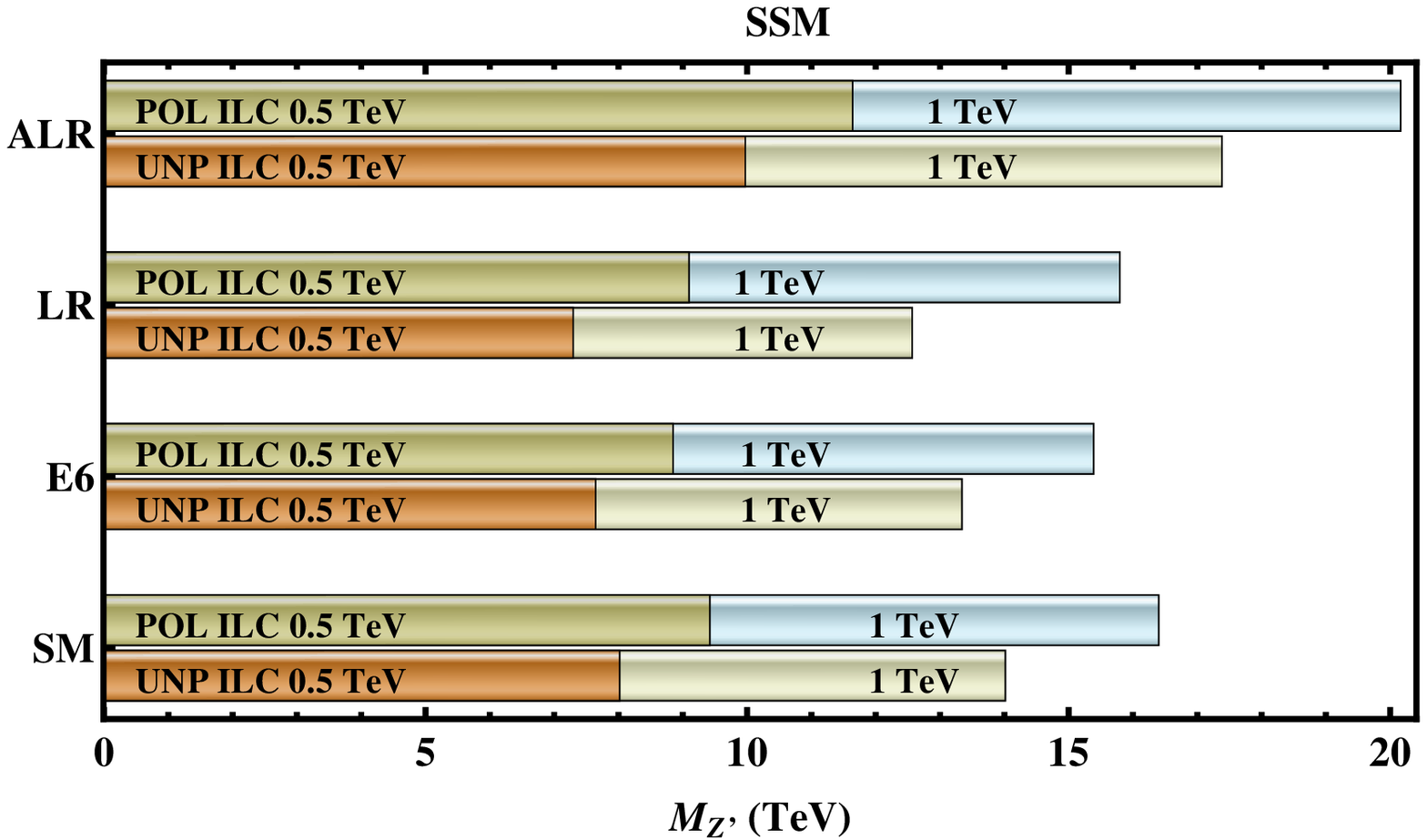}}
\vspace*{0cm} \caption{\label{ID-ALR-SSM} Exclusion reaches on
  $M_{Z'}$ at 95\% C.L. obtained from combination of all processes
  $e^+e^- \to f\bar{f}$ at $\sqrt{s}$=0.5 (1.0) TeV and $\Lumint$=500
  (1000) fb$^{-1}$ in the case when the ALR model (top panel) or SSM
  model (bottom) is assumed to be the `true' models while the others
  are taken as tested models. The unpolarized and polarized cases are
  compared.}
\end{figure}

\par
We now continue the above analysis in a somewhat different direction,
namely, we wish to determine the limiting value of $M_{Z^\prime}$
up to which a particular $\cos\theta$-dependent $E_6$ model, assumed
to be true, can be identified at the ILC in the sense that all
the other, potentially competing, $Z^\prime$ models can be excluded.
The results from this analysis are shown in
Figs.~\ref{ID-E6}--\ref{ID-ALR-SSM}.

\par
Figure~\ref{ID-E6} exhibits the exclusion limits vs.\ $\beta$
on the models LR, ALR, SSM and SM (recall that exclusion
of the SM determines the discovery reaches), once a
$E_6$ model is `true' (all processes combined). For the LR `tested'
model, the corresponding curve in Fig.~\ref{ID-E6} is obtained, for
each $\beta$, by varying $\alpha_{\rm LR}$ in the full allowed
range, which gives LR exclusion limits $M_{Z^\prime}(\alpha_{\rm LR})$,
and choosing the minimum value of such $M_{Z^\prime}$s (in this way
the {\it whole} class of LR models, as well as the LRS, are
excluded). The solid line labelled `SM' represents the
discovery reach, i.e., the $Z^\prime$ mass up to which the
SM can be excluded. The overall identification range is
shown as the shaded (yellow) region. One
can see that, in this case, the identification of the class
of $E_6$ models considered here is basically determined by the
exclusion of the class of the LR models and, for the `central'
values of $\cos\beta$, by the SM (i.e., by the discovery reach).
Consequently the ID-limit
is, in a (somewhat broad) range around $\cos\beta = 0$,
essentially identical to the discovery limit, whereas it is
substantially smaller in the two intervals close
to $\vert\cos\beta\vert = 1$. Figure~\ref{ID-E6} shows that,
numerically, for $\vert\cos\beta\vert<0.9$ the ID-limit is as
large as $M_{Z'}^{\rm ID}\simeq3-4~\text{TeV}$, and for $\cos\beta$ near
$\pm$1 $E_6$ models become more and more difficult to
distinguish from the competitor ones.
The right panel of Fig.~\ref{ID-E6} shows the corresponding
identification reaches for the polarized case,
$\vert P^-\vert=0.8$ and $\vert P^+\vert=0.6$,
and the quantitative improvements that can be achieved
in this case.

Similarly, the identification limits on LR models vs.\ the parameter
$\alpha_{\rm LR}$ can be read off from Fig.~\ref{ID-LR}. The curve
labelled as `$E_6$' is obtained by a procedure analogous to
the curve `LR' in Fig.~\ref{ID-E6}, and the solid curve `SM'
represents the exclusion limits of the SM (hence the discovery
reaches). In this case, the identification of the class of LR
$Z^\prime$s turns out to be determined basically by the
exclusion of the class of $E_6$ models, generally not so much
below the discovery limit for all values of $\alpha_{\rm LR}$.
On the other hand, the figure shows rather high identification
limits, of the order of
$M_{Z'}^{\rm ID}\simeq 3.0-4.6~\text{TeV}$ in the range, say,
$0.9\lsim \alpha_{\rm LR}\lsim\sqrt{2}$, whereas they
substantially decrease for smaller $\alpha_{\rm LR}$.

\begin{table}[h!]
\begin{center}
\label{tab:identify}
\renewcommand{\tabcolsep}{.55em}
\caption{\label{Tab:luminosity}
Required integrated luminosity, $\Lumint [\text{fb}^{-1}]$, at the two
 energies $\sqrt{s}=0.5$ and 1~TeV and with polarized beams, required for
model
 identification. Three mass values, $M_{Z^\prime}=2.5~\text{TeV}$, 3.5~TeV
and
 4.5~TeV, assumed determined at the LHC, are considered.}
\vspace{.175in}
\begin{tabular}{|c|ccc|ccc|ccc|ccc|ccc|}
\hline \multirow{2}{*}{$\sqrt{s}$} & \multicolumn{3}{c|}{E6} &
\multicolumn{3}{c|}{LR}
&  \multicolumn{3}{c|}{ALR} & \multicolumn{3}{c|}{SSM} \\
& 2.5& 3.5& 4.5& 2.5& 3.5& 4.5& 2.5& 3.5& 4.5& 2.5& 3.5& 4.5\\
\hline
0.5~TeV & 49.6&  225& 785& 51.0&  241& 944& 2.3& 9.4& 27.1& 2.1& 8.5& 24.0
\\
1~TeV   &  9.1& 41.6& 125&  9.4& 42.3& 128& 0.4& 2.0&  5.9& 0.4& 1.8& 5.3 \\
\hline
\end{tabular}
\end{center}
\end{table}

We can conclude, from Figs.~\ref{ID-E6} and \ref{ID-LR},
that the identification reach at the ILC, already at
$\sqrt{s}=0.5~\text{TeV}$ and  $\Lumint$=500 fb$^{-1}$,
exceeds the corresponding discovery reach at the LHC. In fact,
the full integrated luminosity considered here might be not quite
indispensible for this identification. In Table~\ref{Tab:luminosity}
we show the required integrated luminosity, at the two ILC energies
of 0.5 and 1~TeV, for the identification of these different models,
realized as a $Z^\prime$ at 2.5, 3.5 or 4.5~TeV (within the
discovery reach of the LHC).

Finally, in Fig.~\ref{ID-ALR-SSM} we summarize the
information, of a similar kind as represented in
Figs.~\ref{ID-E6} and~\ref{ID-LR}, relevant to the cases
where the ALR model or the SSM model is assumed `true'
(upper and lower panels, respectively). As usual, the figure
shows the limiting value of $M_{Z^\prime}$ at which the other
$Z^\prime$ models can be excluded, the SM being one of them.
The two energies 0.5~TeV and 1~TeV are
considered, the different processes (\ref{proc_ff}) are
combined in the $\chi^2$, and the unpolarized and polarized
cases ($|P^-|=0.8$, $|P^+|=0.6$) are compared. The entries
for the LR and $E_6$ models in this figure refer to the worst case,
i.e., similar to the procedures adopted for the
`LR' and `$E_6$' curves in Figs.~\ref{ID-E6}
and~\ref{ID-LR}, adopting the lowest value of
$M_{Z^\prime}$ as $\beta$ and
$\alpha_\text{LR}$ are varied, in order to represent a whole class
of models. Specifically, in Fig.~\ref{ID-ALR-SSM}, one can easily
read off the identification reaches for ALR and SSM models at
0.5~TeV: $M_{Z'_{\rm ALR}}^{\rm ID}$
and $M_{Z'_{\rm SSM}}^{\rm ID}$ are both 8--9~TeV
if polarization is available.  In a sense, the ALR and SSM
models are the most `orthogonal' ones since, if either of them
is assumed `true', the other one can be excluded up to a
really high value of $M_{Z^\prime}$.

\subsection{$Z'$ mass not known}

\begin{figure}[hp] %
\centerline{ \hspace*{-0.0cm}
\includegraphics[width=6.5cm,angle=0]{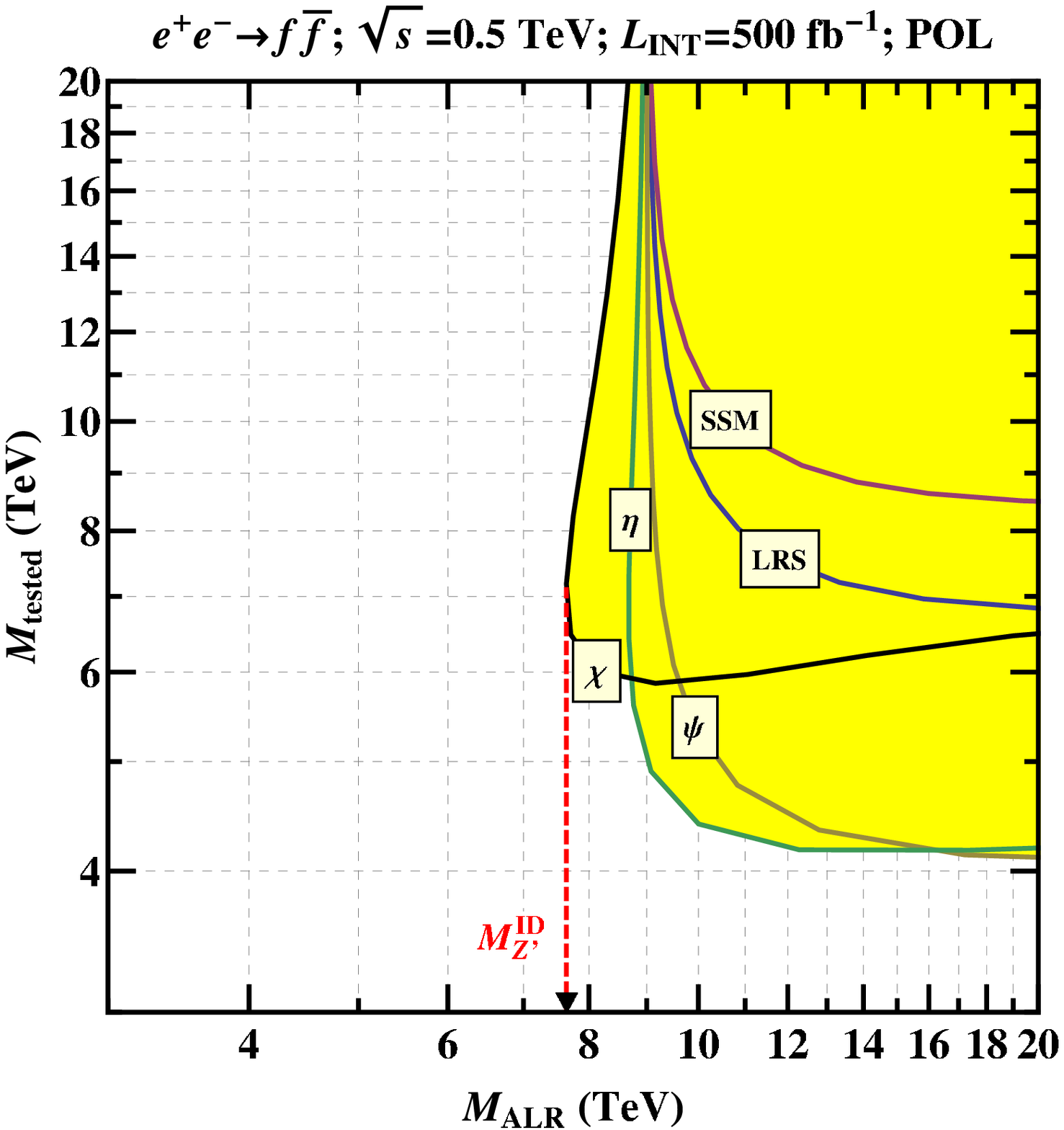}
\hspace*{-0.2cm}
\includegraphics[width=6.5cm,angle=0]{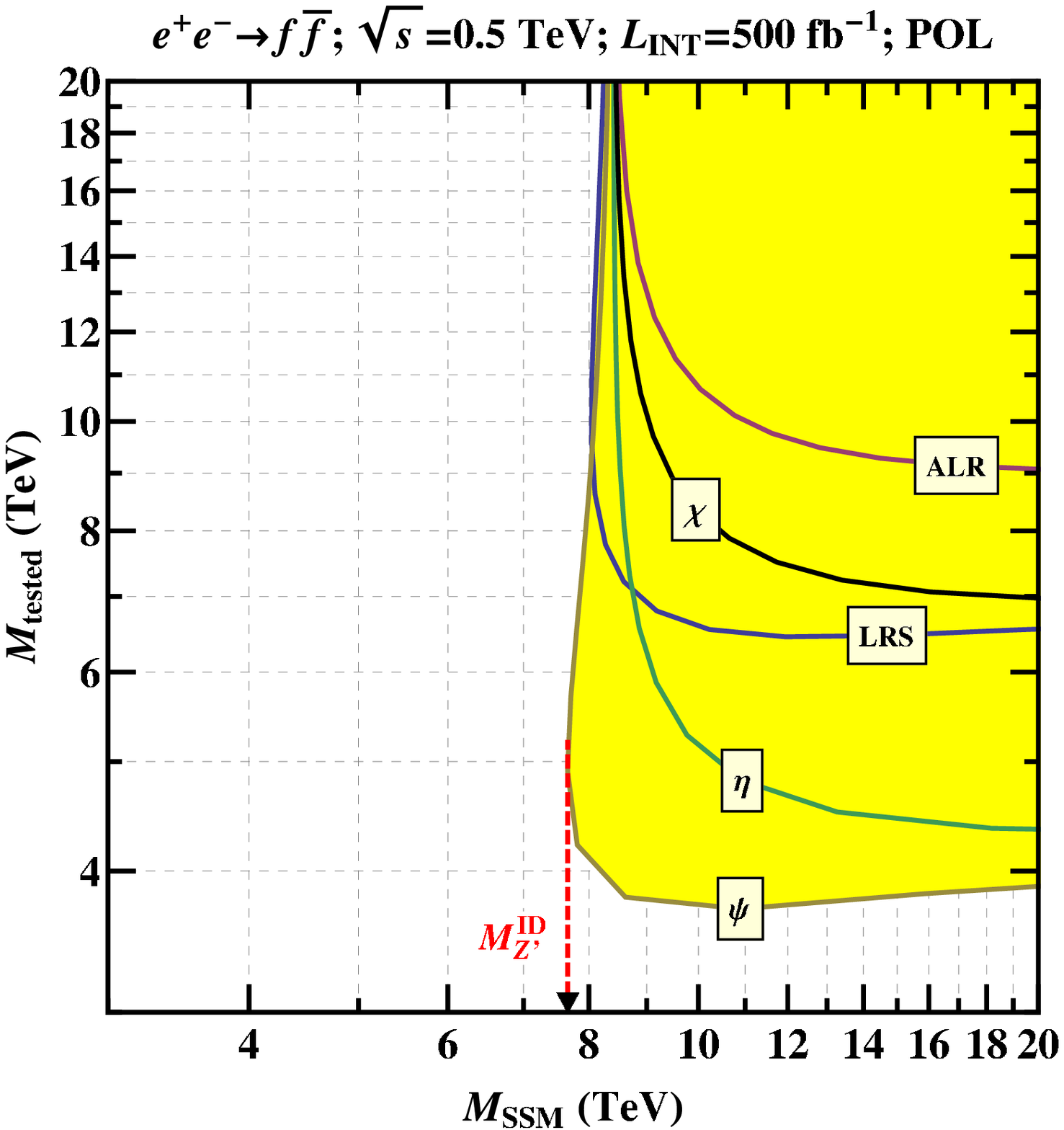}}
\centerline{ \hspace*{-0.2cm}
\includegraphics[width=6.5cm,angle=0]{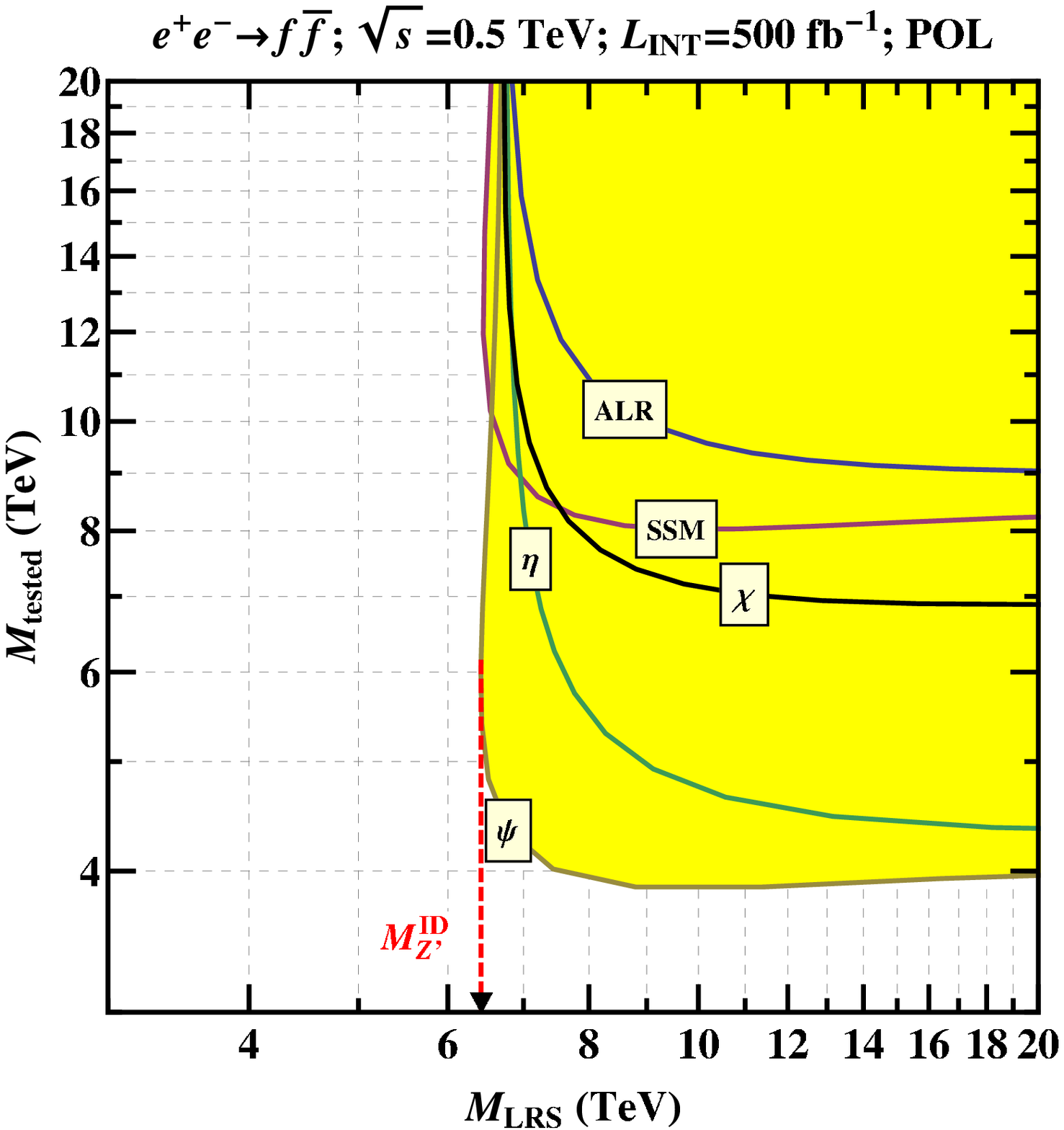}
\hspace*{-0.2cm}
\includegraphics[width=6.5cm,angle=0]{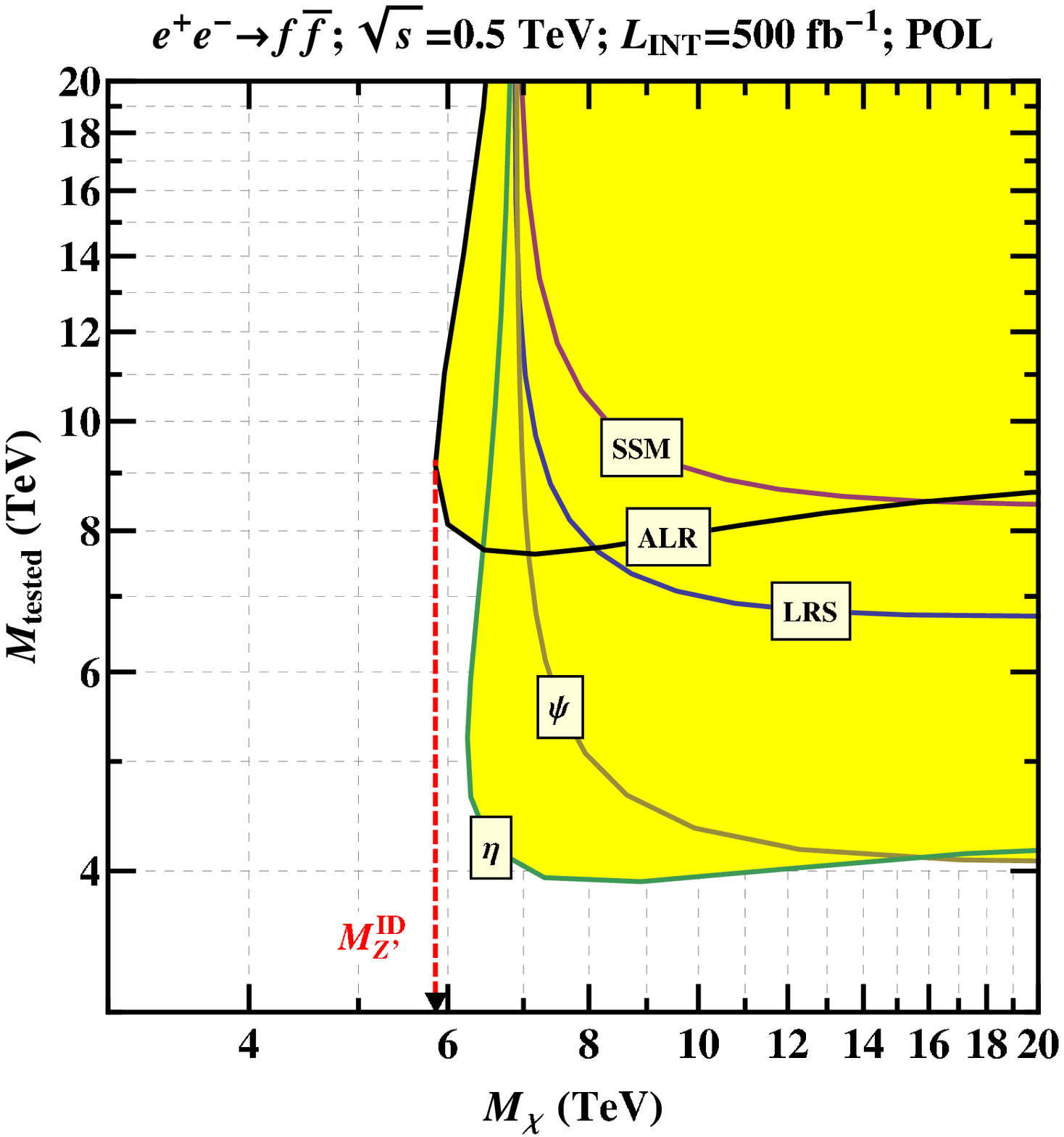}}
\centerline{ \hspace*{-0.2cm}
\includegraphics[width=6.5cm,angle=0]{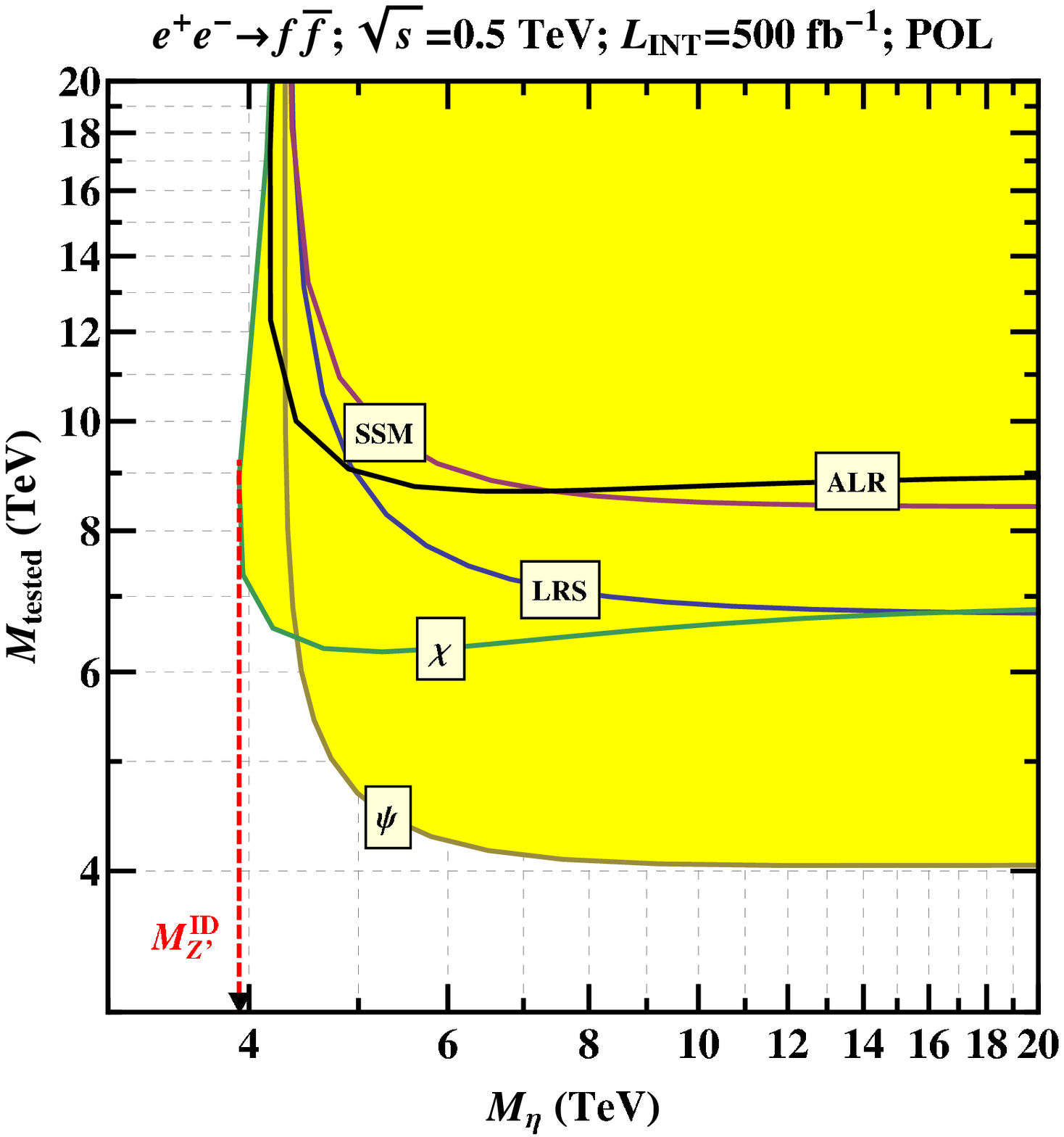}
\hspace*{-0.2cm}
\includegraphics[width=6.5cm,angle=0]{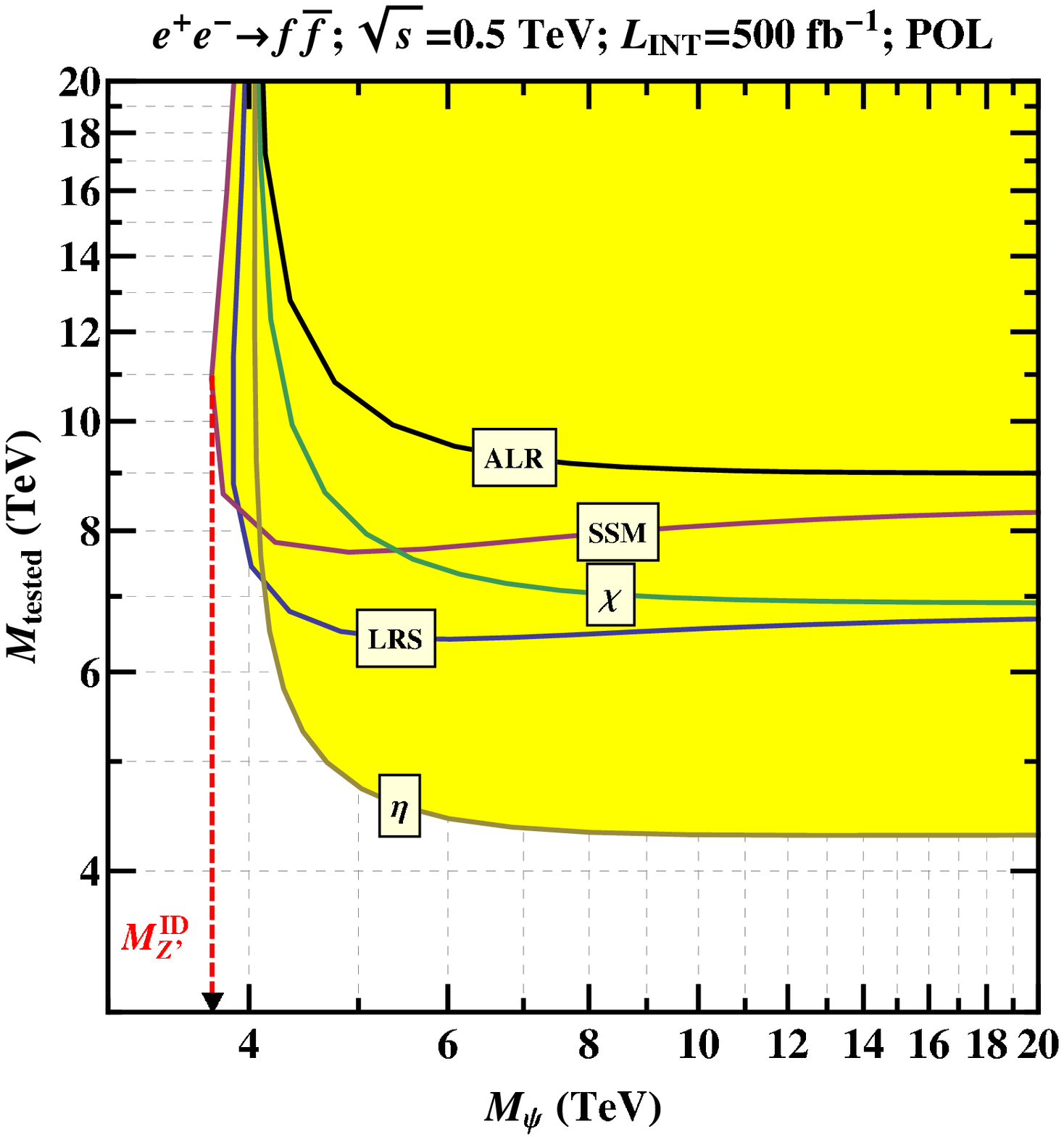}}
\caption{\label{fig-yellow} Regions of confusion (yellow shaded areas)
  between a `true' $Z'$ model and the `tested' $Z'$ models in the mass
  plane ($M_{\rm true}, M_{\rm tested}$) at 95\% C.L.  obtained from
  combination of all polarized processes $e^+e^- \to f\bar{f}$ at
  $\sqrt{s}$=0.5 TeV and $\Lumint$=500 fb$^{-1}$.  The dashed lines indicate the identification reach.}
\end{figure}

The second kind of situation is met in the case
where the $Z'$ mass cannot be known {\it a priori}, e.g.,
the $Z'$ is too heavy to be discovered at the LHC
[say, $M_{Z'}>$4--5~TeV], but deviations from
the SM predictions can still be observed at the ILC.
Actually, models with different $Z^\prime$ masses and
coupling constants can in principle be the source of
a deviation from the SM predictions observed at the ILC.
With the coupling constants held fixed numerically at the
theoretical values pertinent to the $Z^\prime_i$ and
$Z^\prime_j$ models under consideration, the
$\chi_{ij}^2$ of Eq.~(\ref{chisquareprime}) becomes a function of the
two masses, $M_{Z^\prime_i}$ and $M_{Z^\prime_j}$, both assumed to lie
in the respective ILC discovery ranges. In this case, one can derive a
contour in the two-dimensional $(M_{Z^\prime_i},M_{Z^\prime_j})$ plane
where to each value of $M_{Z^\prime_i}$ is associated a value
${\overline M}_{Z^\prime_j}$ such that, for all $M_{Z^\prime_j}>
{\overline M}_{Z^\prime_j}$, the value of $\chi^2_{ij}$ in
(\ref{chisquareprime}) is consistent with `confusion' of $i$ and $j$
at the desired confidence level. The region encircled by such a contour
will be the `confusion' (or `no distinction') domain between the
`true' model $i$ and the `tested' model $j$ and correspondingly, in
the complementary domain the hypothesis $j$ could be {\it excluded} if
$i$ is assumed to be `true'. We refer to this latter, complementary, region
as `resolution' region.

\par
One can iterate this procedure and generate pairwise `confusion' and
`exclusion' regions in the two-dimensional planes of parameters for
all models $j\ne i$. As will be illustrated graphically in the
remaining part of the paper, a common feature of such `exclusion'
regions is that the relevant contours admit, for each $j$ (and
obviously fixed $i$) a minimum value $M_{Z^\prime_i}^{(j)}$ such that,
for any value $M_{Z^\prime_i}< M_{Z^\prime_i}^{(j)}$, the `tested'
model $j$ can be {\it excluded} regardless of $M_{Z^\prime_j}$.  We
finally assume, as identification limit on the $i$ model at ILC,
the smallest of the values $M_{Z^\prime_i}^{(j)} $for $j\ne i$, for
which {\it all} tested models will be excluded by the hypothesis of
$i$ being `true'.  Of course, such ID-value of $M_{Z^\prime}$ should
be smaller (or at most equal), than the ILC discovery reach on model
$i$.  This procedure can finally be iterated, in turn, to all the
different $Z^\prime$ models and the assessment of corresponding
ID-reaches. This naive $\chi^2$ procedure can also be extended in a
straightforward way to estimating exclusion ranges---and corresponding
identification limits---in the cases where $\cos\beta$- and/or
$\alpha_{\rm LR}$-dependent $Z^\prime$ models are considered in
Eq.~(\ref{chisquareprime}).

\par
Examples of pairwise `confusion' regions and corresponding contours,
relevant to the $Z'$ models chosen in Fig.~\ref{discovery}, are shown
in Fig.~\ref{fig-yellow}. In this figure, the various steps of the
procedure outlined above, as well as the final derivation of the
ID-limits, can easily be followed.  As an example of how to read this
figure, consider the hypothesis that the $\eta$ model is `true' (lower
left panel), with $M_{Z^\prime}=6~\text{TeV}$. Then, if instead the
$\psi$ or $\chi$ model should be true, the mass would have to exceed
4.2 or 6.3~TeV, respectively.

\begin{figure}[htb]
\vspace*{0.0cm} \centerline{ \hspace*{-2.0cm}
\includegraphics[width=10.5cm,angle=0]{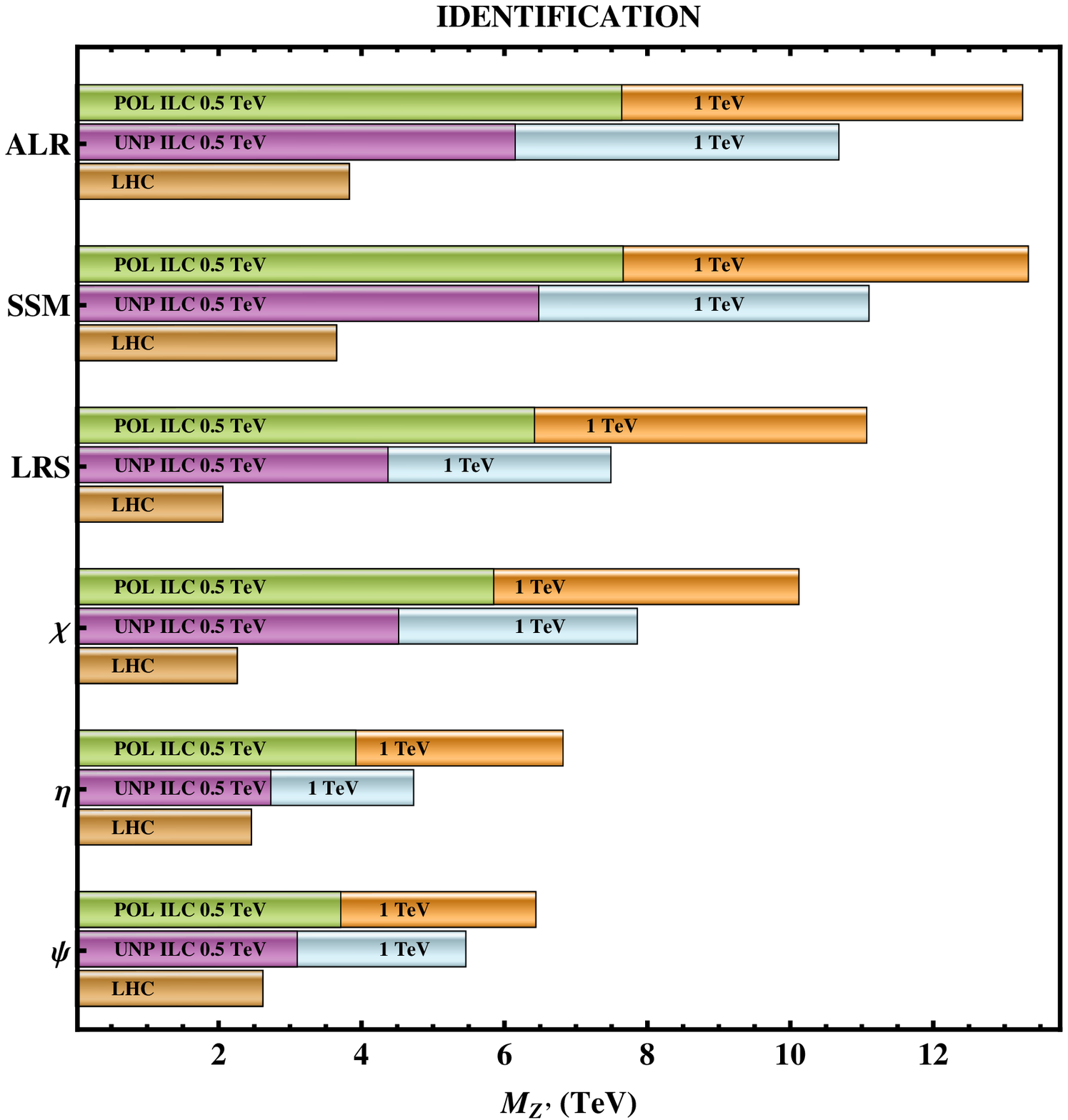}}
\vspace*{0.cm} \caption{\label{identification} Comparison of the
$Z^\prime$-model distinction bounds on $M_{Z'}$ obtained from
combined analysis of the unpolarized and polarized processes
(\ref{proc_ff})  at the ILC with $\sqrt{s}=0.5$~TeV (1 TeV) and
$\Lumint=500$ fb$^{-1}$ (1000 fb$^{-1}$), compared to the results
expected from Drell-Yan processes at the LHC at 95\% C.L.
\cite{Osland:2009tn}.  Two options of polarization are considered:
unpolarized beams $P^-= P^+=0$ and  both beams are polarized,
$\vert P^-\vert=0.8$ and $\vert P^+\vert=0.6$. }
\end{figure}

\par
Finally, Fig.~\ref{identification} shows the comparison of
identification reaches or distinction bounds on the $Z^\prime$-models
considered in Fig.~\ref{discovery}, together with the corresponding
bounds on $M_{Z'}$ obtained from the process $pp\to l^+l^- +X$ at the
LHC with c.m.\ energy 14 TeV and time-integrated luminosity 100 ${\rm
  fb}^{-1}$. We assume, for the ILC, the same c.m. energy, luminosty
and beam polarization as in Fig.~\ref{discovery}. The figure speaks
for itself, and in particular clearly exhibits the roles of the ILC
parameters. In
summary, one might be able to distinguish among the considered
$Z^\prime$ models at 95\% C.L. up to $M_{Z^\prime}\simeq 3.1$~TeV (4.0
TeV) for unpolarized (polarized) beams at the ILC (0.5 TeV) and 5.3~TeV
(7.0 TeV) at the ILC (1 TeV), respectively. In
particular, the figure explicitly manifests the substantial role of
electron beam polarization in sharpening the identification
reaches. Positron polarization can also give a considerable
enhancement in this regard (if measurable with the same high accuracy
as for electron polarization), although to a more limited extent in
some cases.

\par
Clearly, our analysis is greatly simplified by the fact that the
vector and axial vector couplings of the considered $Z^\prime$s
are fixed theoretically. If we wanted to determine them in
general, namely, with both masses {\it and} coupling constants {\it a
priori} free variables, the $\chi^2$ analysis should be
five-dimensional with, in addition, the limitation that for
$M_{Z^\prime}\gg\sqrt{s}$ (contact-interaction regime),
$M_{Z^\prime}$ could not be simultaneously
extracted.
In principle, data at different collider energies could be
utilized in this regard, for $Z^\prime$ masses not too far from
$\sqrt{s}$ \cite{Rizzo:1996rx}.

\section{Concluding remarks}
\label{sect:concl}

We have explored in some detail how the $Z^\prime$ discovery reach at
the ILC depends on the c.m.\ energy, on the available polarization, as
well as on the model actually realized in Nature. The lower part of
this range, up to $M_{Z'}\simeq 5$~TeV, will also be covered by the
LHC, but the identification reach at the LHC is only up to
$M_{Z'}<2.2$ TeV.

In this LHC discovery range, the cleaner ILC environment, together
with the availability of beam polarization, allow for an
identification of the particular $Z^\prime$ version
realized. Actually, this ILC identification range extends considerably
beyond the LHC discovery range.  Specifically, the ILC with polarized
beams at $\sqrt{s}=0.5$ TeV and 1 TeV allows to identify all considered
$Z'$ bosons if $M_{Z'}\lsim (6-7)\times\sqrt{s}$. This represents a substantial
extension of the the LHC reach.

\vspace{0.5cm} \leftline{\bf Acknowledgements}
\par\noindent
It is a great pleasure to thank Nello Paver for his contributions to this study,
in particular his critical and constructive comments.
This research has been partially supported by the Abdus Salam
ICTP and the Belarusian Republican Foundation for Fundamental
Research. The work of PO has been supported by the Research
Council of Norway.


\end{document}